\documentclass{article} % For LaTeX2e
\usepackage{iclr2027_conference,times}

\usepackage{amsmath,amsfonts,bm}

\def\eqref#1{equation~\ref{#1}}
\def\1{\bm{1}}

\DeclareMathAlphabet{\mathsfit}{\encodingdefault}{\sfdefault}{m}{sl}
\SetMathAlphabet{\mathsfit}{bold}{\encodingdefault}{\sfdefault}{bx}{n}

\iclrfinalcopy
\usepackage{hyperref}
\usepackage{url}
\usepackage{float}
\usepackage{subcaption}
\usepackage[table,xcdraw]{xcolor}
\usepackage{booktabs}
\usepackage{xspace}
\usepackage{multirow}
\usepackage{enumitem}
\usepackage{graphicx}
\usepackage{wrapfig}
\usepackage{tcolorbox}
\tcbuselibrary{listings,breakable}
\usepackage[normalem]{ulem}
\usepackage{titletoc}

\definecolor{appendixpurple}{HTML}{7040A0}
\definecolor{grpA}{HTML}{FBE9EC}   % Direct Reasoning
\definecolor{grpB}{HTML}{FDF0DC}   % Teacher
\definecolor{grpC}{HTML}{EAF0FA}   % Student
\definecolor{ourrow}{HTML}{F2F2F2} % 我们的方法行
\definecolor{citeblue}{HTML}{5588CB} % Cite Color
\definecolor{promptbg}{HTML}{EAF2FC}
\definecolor{prompttitle}{HTML}{234B79}
\hypersetup{
    colorlinks=true,
    linkcolor=citeblue,
    citecolor=citeblue,
    urlcolor=citeblue
}

\newcommand{\ours}{\texttt{IG-OPD}\xspace}

\title{Information-Gain Calibrated On-Policy Distillation for Deep Search Agents}

\author{%
Chenxi Li$^{1*}$, Xueyang Feng$^{2*}$, Songyang Liu$^{3}$, Chenxu Wang$^{3}$, Xiaoying Tang$^{1}$, \\
\textbf{Chaozhuo Li$^{4\dagger}$, Litian Zhang$^{3}$} \\
$^{1}$The Chinese University of Hong Kong, Shenzhen, 
$^{2}$ Renmin University of China, \\
$^{3}$Beijing University of Posts and Telecommunications, \\
$^{4}$Beijing Academy of Artificial Intelligence (BAAI) \\
\texttt{ChenxiLi\_lcx@outlook.com}, 
\texttt{lichaozhuo1991@gmail.com}
}
\begin{document}

\maketitle

\begingroup
\renewcommand{\thefootnote}{}
\footnotetext{% 
Work done during an internship at BAAI.
$^{*}$Equal contribution.
$^{\dagger}$Corresponding author.\\
}
\endgroup

\vspace{-10pt}
\begin{abstract}
On-policy distillation (OPD) has recently emerged as a promising approach for transferring the capabilities of large agents to compact students. In deep search, where agents iteratively retrieve external evidence, a teacher's preference among plausible queries before retrieval may not predict which queries will return useful evidence. Standard OPD nevertheless supervises each student query from the preceding context alone, inducing a failure mode we term \textbf{\textit{evidence-blind supervision (EBS)}}. To understand this, we regroup the OPD objective by interaction turns and reveal two implicit priors: a \textit{budget prior} that allocates supervision by token volume, and a \textit{direction prior} that favors tokens preferred by the teacher, neither of which accounts for the evidence returned by the current query. Our empirical analysis shows that both priors contribute to EBS in practice: the first search turn accounts for 51.1\% of total absolute information gain but receives only 18.0\% of the supervision budget, while 43.7\% of search queries have teacher rewards that disagree in sign with post-retrieval information utility. To mitigate this, we propose \ours, which uses post-retrieval information gain (IG) to calibrate both priors: reallocating supervision toward information-critical turns while calibrating query preferences using the utility of retrieved evidence. Experiments on six multimodal search benchmarks show that \ours improves average accuracy by 3.4 and 1.7 percentage points for 2B and 4B students over standard OPD, enabling the 4B student to surpass its 8B teacher on MMSearch and BrowseComp-VL.
\end{abstract}

\section{Introduction}
Real-world queries often require knowledge beyond a model's pretrained parametric memory, such as time-sensitive facts and long-tailed entities~\citep{lewis2020retrieval,mallen2023not,jiang2024mmsearch}. To address this, search agents equip Large Multimodal Models (LMMs) with external search tools to iteratively reason over retrieved evidence. However, acquiring these capabilities typically requires large-scale agentic reinforcement learning (RL) on heavyweight models (e.g., MMSearch-R1~\citep{wu2025mmsearch}, OpenSearch-VL~\citep{chen2026opensearch}, and HyperEyes~\citep{li2026hypereyes}), making the agents expensive to train and deploy. Thus, on-policy distillation (OPD) offers a natural route to transfer such RL-acquired capabilities to compact students by supervising the student's own rollouts with dense teacher signals derived from a reverse-KL objective. Previous work on OPD focused mainly on single-turn settings such as mathematics and coding, emphasizing stabilizing training under high-variance rewards~\citep{hao2026stabilizing,xie2026trust,jin2026entropy} or combining OPD with outcome-based RL~\citep{tan2026atod,xu2025kdrl}. Recent work has extended OPD to multi-turn settings, such as embodied reasoning and web navigation, focusing primarily on mitigating error accumulation across interaction turns~\citep{wang2026tcod,li2026policy,zhou2026sage}.

\begin{figure*}[t]
  \centering
  \includegraphics[width=\textwidth]{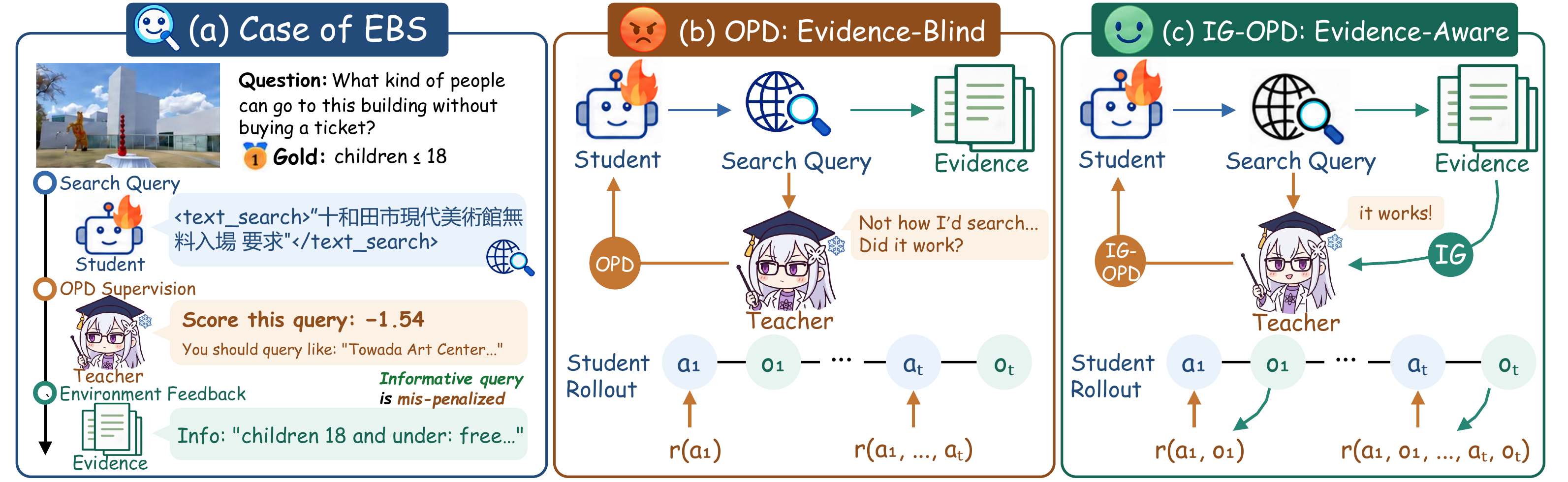}
  \caption{Illustration of \textbf{evidence-blind supervision} in OPD for search agents. \textbf{(a) EBS example} shows that a useful search query is mis-penalized even though it subsequently retrieves the gold evidence. \textbf{(b) Standard OPD} supervises the agent's actions without considering the returned evidence. \textbf{(c) IG-OPD} enables evidence-aware supervision by incorporating post-retrieval information gain.}
  \vspace{-10pt}
  \label{fig:motivation}
\end{figure*}

Despite their success, these approaches typically derive their supervision of each student action using the preceding context alone, \textit{without} accounting for the subsequent environmental feedback. This consequence-agnostic supervision becomes problematic in deep search, where we observe a recurring failure mode: a search question may admit several plausible queries that differ in wording, language, or retrieval path, but their usefulness still depends on the evidence they actually retrieve. Even a query preferred by the teacher might yield little useful evidence, while a less favored one could retrieve the evidence needed to answer the question. Yet standard OPD still scores the student's query using teacher–student token likelihoods conditioned only on the pre-retrieval context, so the resulting score may not reflect the query’s realized utility. It can therefore penalize a useful query simply because the teacher preferred a different one. Figure~\ref{fig:motivation} illustrates a case: the student correctly identifies the ``Towada Art Center'' and issues a Japanese query about free admission. Standard OPD still scores it at $-1.54$, favoring the teacher’s wording even though the teacher’s query retrieves no useful evidence. The student’s query, by contrast, retrieves the gold evidence, yet the supervision ignores it. We refer to this failure mode as \textbf{evidence-blind supervision (EBS)}.

To understand this issue, we regroup the OPD objective by interaction turns, revealing two implicit priors that govern supervision allocation and update direction (see Appendix~\ref{app:opd_decomposition} for the derivation):
\vspace{-3pt}
{\small
\begin{equation}
J_{\mathrm{OPD}}
= \mathbb{E}_{\tau \sim \pi_S}\!
\sum_{t=1}^{T}
\underbrace{\frac{n_t}{N}}_{\text{budget prior}}
\cdot
\frac{1}{n_t}\sum_{i \in \mathrm{turn}_t}\underbrace{ \log\frac{\pi_T}
{\pi_S}}_{\text{direction prior}}
\label{eq:opd_decomposition}
\end{equation}
}
\vspace{-14pt}

where $n_t$ is the token count of turn $t$, $N=\sum_t n_t$ is the trajectory length, and the log-ratio between teacher $\pi_T$ and student $\pi_S$ is evaluated at each student-sampled token. Under this formulation, (i)~the \textit{budget prior} $n_t/N$ allocates each turn's share of supervision by raw token volume, so a verbose synthesis turn can outweigh a concise search turn even when the latter contributes more to finding the answer. Our empirical analysis in Section~\ref{sec:ig_ebs} confirms this effect: the first search turn accounts for 51.1\% of the total information criticality, yet receives only 18.0\% of OPD's token budget. Meanwhile, (ii)~the \textit{direction prior} $\log\frac{\pi_T}{\pi_S}$ favors queries closer to the teacher's preference, regardless of whether they retrieve useful evidence. Empirically, we find that roughly 43.7\% of query supervision is inconsistent with post-retrieval information utility. Together, these limitations naturally motivate a key question: \textit{\uline{How can we seamlessly integrate information available from post-retrieval evidence into the OPD process without distorting or destabilizing training dynamics?}}

Building upon these, we introduce information gain (IG) as a lightweight post-retrieval signal to mitigate the two priors in EBS. Therefore, we propose \textbf{I}nformation-\textbf{G}ain Calibrated \textbf{O}n-\textbf{P}olicy \textbf{D}istillation (\ours), a novel OPD framework for evidence-aware supervision. Specifically, \ours employs Information-Guided Turn-Credit Calibration (ITC) to adjust the \textit{budget prior} by information criticality, and Information-Guided Query-Preference Calibration (IQC) to reweight the \textit{direction prior} by information utility. Extensive experiments in a live web environment demonstrate that \ours improves average accuracy by +3.4 and +1.7 for 2B and 4B students, respectively. It also boosts the gold-evidence hit rate by 14.7\% relatively, with the 4B student surpassing its 8B teacher on MMSearch and BrowseComp-VL. To summarize, our main contributions are as follows:
\begin{itemize}[leftmargin=1.5em]
    \item We identify and formalize \textbf{EBS}, a systematic failure mode of OPD in deep search agents where supervision is assigned without considering retrieved evidence, showing how supervision assigned before retrieval can under-credit information-critical turns and penalize useful queries. 
    \item We propose \ours, an evidence-aware OPD framework that uses post-retrieval information to calibrate OPD's two priors through complementary mechanisms: \textbf{ITC} reallocates cross-turn budgets by information criticality, while \textbf{IQC} calibrates query preferences with evidence utility.
    \item Across six multimodal search benchmarks, \ours improves average accuracy over OPD by 3.4 and 1.7 percentage points for the 2B and 4B students, respectively. The 4B student also surpasses its 8B teacher on MMSearch and BrowseComp-VL, with zero inference overhead.
\end{itemize}
\vspace{-2pt}

\section{Related Work}
\label{related_works}
\vspace{-5pt}
\textbf{Multimodal search agents.} Multimodal search agents extend question answering beyond parametric memory by retrieving related web evidence. Recent systems mainly use reinforcement learning (RL) to optimize multi-step retrieval and reasoning~\citep{zhang2025deep,lin2025comprehensive,hong2025deepeyesv2}. For example, MMSearch-R1 and WebWatcher train search policies with GRPO~\citep{wu2025mmsearch,geng2026webwatcher}, while OpenSearch-VL combines multi-hop training data with safeguards against cascading tool errors~\citep{chen2026opensearch}. However, sparse outcome rewards complicate credit assignment and make training costly, motivating more sample-efficient supervision.

\textbf{On-policy distillation.} On-policy distillation (OPD) provides dense teacher feedback on student-generated trajectories, offering a sample-efficient approach to capability transfer. Existing studies mainly address three vulnerabilities in OPD. \textbf{Training instability} arises from high-variance distillation rewards, motivating reward normalization~\citep{hao2026stabilizing}, reward clipping~\citep{ko2026scaling}, and proximal teacher distributions~\citep{xie2026trust}. \textbf{Indiscriminate imitation} overlooks differences in uncertainty and supervision quality across tokens and trajectories. To solve this, related methods introduce entropy-aware objectives~\citep{jin2026entropy}, token selection~\citep{huang2025selectkd,xu2026tip}, or selective filtering and reweighting~\citep{li2026filter,zheng2026scope}. In multi-turn agents, \textbf{compounding errors and distribution shift} further undermine teacher guidance as early mistakes alter subsequent interactions~\citep{zhou2026turnopd,liao2026multi}. TCOD progressively extends the distillation horizon~\citep{wang2026tcod}, while teacher intervention steers trajectories toward more reliable guidance~\citep{li2026policy}. SOD and SAGE instead adapt supervision across steps or turns to mitigate unreliable signals on diverging histories~\citep{zhong2026sod,zhou2026sage}. Besides, other approaches \textbf{combine distillation with RL objectives}~\citep{xu2025kdrl,xu2026rlkd,zhang2026reinforcement}, with ATOD gradually shifting from OPD to RL~\citep{tan2026atod}.

While these strategies refine supervision delivery from various angles, they fundamentally suffer from evidence blindness in multi-hop search: supervising student actions without considering the post-retrieval utility of generated queries, ultimately giving rise to EBS. To overcome this, \ours leverages information gain as a post-retrieval signal to calibrate both turn budgets and update directions, effectively rectifying the misalignment between OPD priors and realized evidence utility.

\vspace{-5pt}
\section{Preliminaries}
\label{preliminaries}
\vspace{-8pt}

\subsection{Problem Formulation: Multi-Turn Search Agents}
\vspace{-4pt}
Following \citet{wang2026information}, we consider a dataset $\mathcal{D}=\{(x,y^\star)\}$, where each query $x=(q,v)$ pairs a textual question $q$ with an optional image $v$, and $y^\star$ is the gold answer. Starting from $x$, the agent iteratively alternates between internal deliberation and tool invocation (e.g., \texttt{text\_search}, \texttt{crop}) until concluding with a final response, yielding a multi-turn rollout $\tau=(\tau_1,\dots,\tau_T)$. Each intermediate turn $\tau_t=(h_t,a_t,o_t)$ ($t<T$) encapsulates a chain of thought $h_t$ in \texttt{<think>}, a tool call $a_t$ in \texttt{<tool\_call>}, and the resulting evidence $o_t$. For the final turn $\tau_T=(h_T,\hat{y})$, the search agent will organize and reason over collected evidence to produce a final synthesis response $\hat{y}$ wrapped in \texttt{<response>}. Because observations are externally produced by the environment, training losses are computed exclusively on tokens generated by the student policy. To better formalize \ours in the next section, we denote $H_t$ as the cumulative interaction history after turn $t$, initialized with $H_0=x$. Moreover, we use $\mathcal{I}_t$ to index the generated tokens in turn $t$, with $n_t=|\mathcal{I}_t|$ and total trajectory length $N=\sum_{t=1}^{T}n_t$. For each token $i\in\mathcal{I}_t$, $u_{t,i}$ denotes the sampled token and $z_{t,i}$ its complete conditioning prefix, which incorporates all preceding context and observations.
\vspace{-8pt}
\subsection{On-Policy Distillation}
\label{sec:opd_preliminary}
\vspace{-3pt}
On-policy distillation (OPD) transfers knowledge from a frozen teacher $\pi_T$ to a student $\pi_S$ with parameters $\theta$ on the student's own rollouts $\tau\sim\pi_S(\cdot\mid x)$. Here, the teacher supervises on the student's state distribution rather than on fixed demonstrations, thereby eliminating the exposure gap of offline distillation. The objective of OPD is the reverse KL divergence $D_{\mathrm{KL}}\big(\pi_S(\cdot\mid z_{t,i})\,\|\,\pi_T(\cdot\mid z_{t,i})\big)$ averaged over student-generated tokens. Following standard practice in large-scale agent distillation~\citep{wang2026tcod}, we instantiate the objective using the sampled token log-likelihood ratio:
\begin{equation}
\widehat{\mathcal{L}}_{\mathrm{OPD}}(\theta)
=-\,\mathbb{E}_{x\sim\mathcal{D},\,\tau\sim\pi_S}
\Big[\frac{1}{N}\sum_{t=1}^{T}\sum_{i\in\mathcal{I}_t}
D^{\mathrm{OPD}}_{t,i}\Big],
\qquad
D^{\mathrm{OPD}}_{t,i}
\equiv
\log\frac{\pi_T(u_{t,i}\mid z_{t,i})}{\pi_S(u_{t,i}\mid z_{t,i})}.
\label{eq:sampled_opd}
\end{equation}
Regrouping Eq.~\ref{eq:sampled_opd} by interaction turns recovers the objective decomposition in Eq.~\ref{eq:opd_decomposition}, revealing two implicit priors. Crucially, $D^{\mathrm{OPD}}_{t,i}$ depends solely on the prefix $z_{t,i}$ and never consults the observation $o_t$ returned by turn $t$, giving rise to the failure mode of EBS. To mitigate this, we show how to leverage realized post-retrieval evidence to dynamically calibrate this OPD objective in Section~\ref{method}.

\vspace{-8pt}
\section{Information-Gain Calibrated On-Policy Distillation}
\label{method}
\vspace{-5pt}

Eq.~\ref{eq:opd_decomposition} shows that vanilla OPD determines both \textit{how much} each turn is supervised ($n_t/N$) and \textit{in which direction} tokens are pushed ($\log\frac{\pi_T}{\pi_S}$) via quantities fixed before search execution. \ours preserves the dense teacher signal while calibrating both priors with a posterior signal available upon observation return. We first define this post-retrieval information gain $g_t$ and elaborate on the practical effects of EBS (Sec.~\ref{sec:ig_ebs}), then recalibrate turn supervision budgets (ITC, Sec.~\ref{sec:itc}) and adjust query update directions (IQC, Sec.~\ref{sec:iqc}). Finally, Sec.~\ref{sec:objective} synthesizes the overall objective.

\vspace{-5pt}
\subsection{Diagnosing Evidence-Blind Supervision with Information Gain}
\label{sec:ig_ebs}
\textbf{Post-retrieval information gain.} We quantify the value of a search turn by how much its evidence changes belief about the gold answer, following IGPO~\citep{wang2026information}. Given the interaction history $H_t$ after the $t$-th observation, we compute length-normalized log-likelihood of the reference answer $y^\star=(y^\star_1,\dots,y^\star_L)$ under the teacher and convert it into a stabilized confidence potential:
\vspace{-5pt}
\begin{equation}
\bar{\ell}_t=\frac{1}{L}\sum_{j=1}^{L}\log\pi_T\!\left(y^\star_j\mid H_t,y^\star_{<j}\right),
\qquad
\Phi_t=\log\!\left(\exp(\bar{\ell}_t)+\epsilon\right),
\label{eq:potential}
\end{equation}
\vspace{-12pt}

where $\epsilon=10^{-3}$ prevents numerical instability in the logarithm as token probabilities approach zero. The information gain of turn $t$ is defined as the potential change induced by its observation: $g_t=\Phi_t-\Phi_{t-1}$ for $t<T$, with $\Phi_0$ evaluated on the initial query $H_0=x$. Intuitively, a higher $g_t$ indicates that the evidence conveys more valuable information. Moreover, these turn-level gains telescope to the net confidence change, anchoring the calibration against collapse~\citep{ng1999policy}.

\begin{wrapfigure}{r}{0.36\textwidth}
  \centering
  % 第一张图（Top）
  \vspace{-15pt}
  \includegraphics[width=\linewidth]{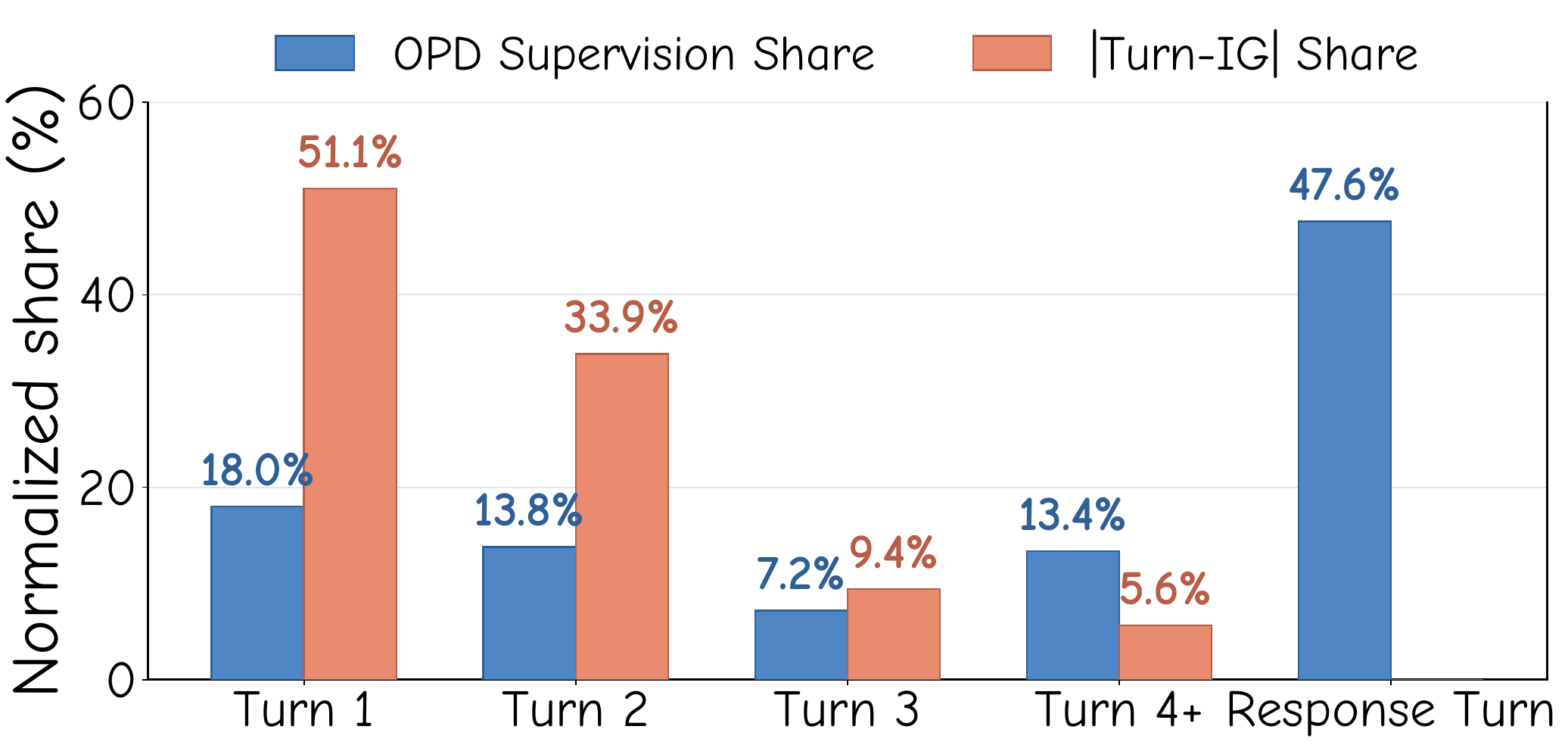}
  \vspace{-4pt} % 压紧两图之间的垂直间距
  % 第二张图（Bottom）
  \includegraphics[width=\linewidth]{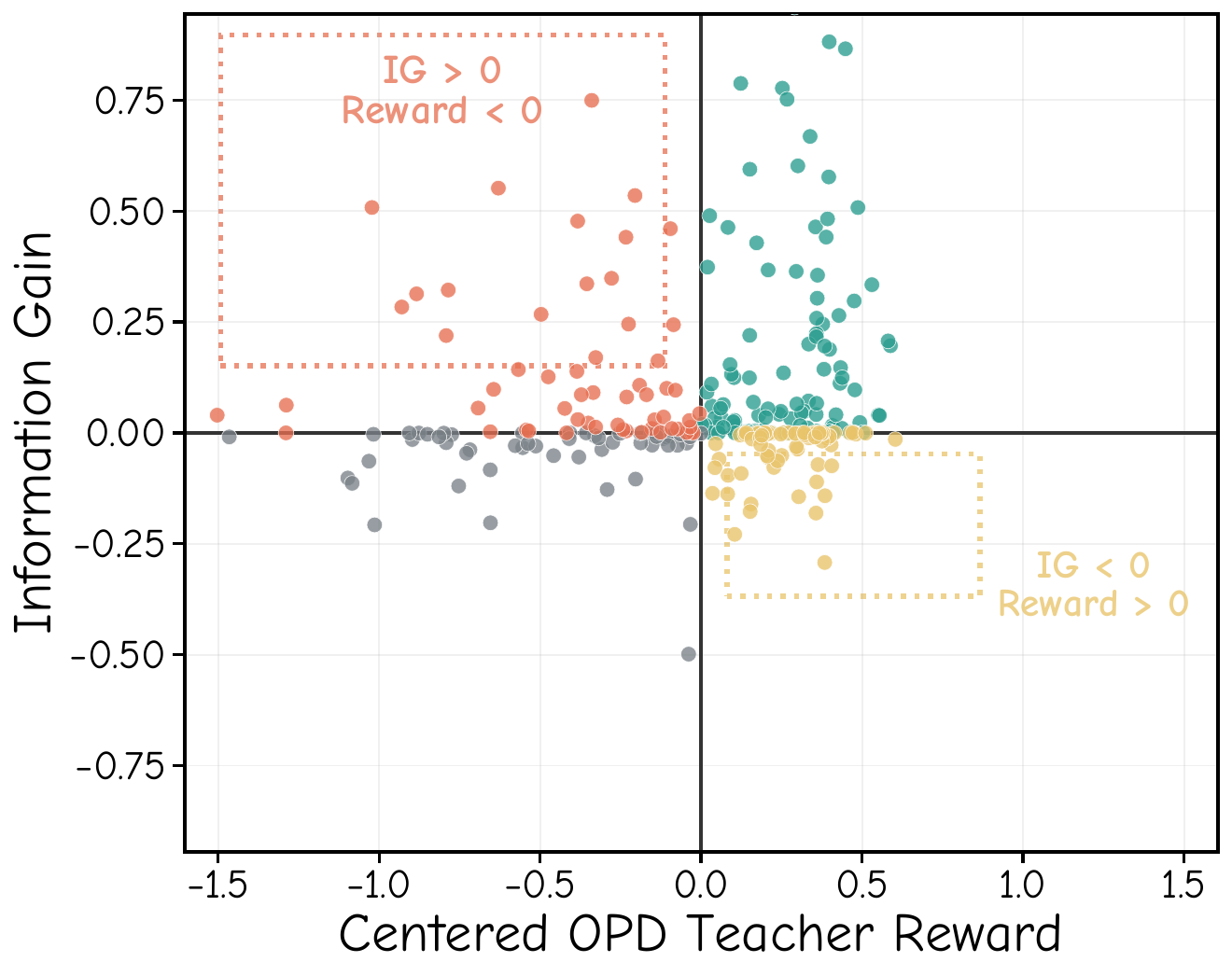}
  \vspace{-10pt} % 压紧底图与 caption 的间距
  \caption{Empirical analysis of EBS under standard OPD's budget prior (Top) and direction prior (Bottom).}
  \label{fig:ebs_diagnosis}
  \vspace{-12pt}
\end{wrapfigure}
\textbf{Empirical Diagnosis of EBS.} Standard OPD assigns its supervision before evidence arrives, naturally prompting us to ask: \textit{How does EBS manifest in practice, and how severely do the two priors mismatch the posterior information utility?} To examine this, we roll out a student agent (Qwen3-VL-2B-Instruct~\citep{bai2025qwen3}) on the MMSearch benchmark~\citep{jiang2024mmsearch}, collecting 300 multi-turn trajectories spanning roughly 1,300 interaction turns. For each turn, we compute both the dense reverse-KL OPD reward scored by the teacher (OpenSearch-VL-8B~\citep{chen2026opensearch}) and its post-retrieval information gain $g_t$. Notably, we centralize the OPD rewards by subtracting their mean across candidate rollouts to offset the negative bias inherent to sampled RKL.

To assess the \textit{budget prior}, we treat each turn's magnitude $|g_t|$ as its information criticality and compare its normalized share against the default budget ratio $n_t/N$, thus revealing a \textbf{Turn-level EBS}. As shown in Figure~\ref{fig:ebs_diagnosis} (Top), the first search turn accounts for 51.1\% of the total $|g_t|$ on average, yet receives only 18.0\% of the supervision budget, yielding just $0.35$ units of supervision per unit of information. In contrast, the final response turn, which primarily synthesizes and organizes previously collected evidence, absorbs 47.6\% of the budget despite contributing zero information gain. These imply that informative turns can suffer from supervision starvation, while verbose turns dominate update gradient.

To assess the \textit{direction prior}, we treat each query's realized signed $g_t$ as an \textbf{empirical} post-retrieval information utility and contrast it with the centered OPD reward, uncovering a severe \textbf{Query-level EBS}. As illustrated in Figure~\ref{fig:ebs_diagnosis} (Bottom), roughly 43.7\% of the search turns fall into the second and fourth quadrants, where the prior OPD reward and post-retrieval $g_t$ disagree in sign. Specifically, 19.8\% of search queries retrieve useful evidence ($g_t > 0$) yet are penalized, while 23.9\% retrieve misleading evidence ($g_t < 0$) yet are rewarded. These results show that \textbf{inherent} teacher preference prior to retrieval is not a faithful proxy for actual evidence utility, causing beneficial search actions to be penalized while misleading ones are falsely reinforced.

These empirical findings naturally motivate us to use IG as a post-retrieval signal and propose \ours to address each symptom in turn, using $|g_t|$ to recalibrate the budget (ITC) and the sign of $g_t$ to recalibrate the direction (IQC). We will also discuss the reliability of IG $g_t$ in Section~\ref{sec:schedule}.

% \begin{figure*}[t]
%     \centering
%     \includegraphics[width=\textwidth]{figs/IG_OPD_method.pdf}
%     \caption{\textbf{Overview of IG-OPD.} After retrieval, ITC redistributes supervision across turns using information-gain magnitude, while IQC adjusts action-critical token scores using signed gain.}
%     \vspace{-10pt}
%     \label{fig:ig_opd_method}
% \end{figure*}

\vspace{-5pt}
\subsection{Information-Guided Turn-Credit Calibration}
\label{sec:itc}
The budget prior $n_t/N$ in Eq.~\ref{eq:opd_decomposition} implicitly assumes the longest turn to dominate the distillation. In deep search,  the final response turn, which merely consolidates collected evidence, is typically far longer than the search turns that acquired it. Consequently, the critical search behavior receives the least supervision, dictated by token volume rather than information value. ITC replaces this length-induced weight with an information-induced one by first normalizing each turn to remove token-length bias, and then modulating it using the turn's posterior information magnitude:
\begin{equation}
q_t=\frac{1+\lambda\tanh\!\left(|g_t|/\sigma_g\right)}{\sum_{j=1}^{T}\left[1+\lambda\tanh\!\left(|g_j|/\sigma_g\right)\right]},
\qquad \sum_{t=1}^{T}q_t=1,
\label{eq:itc_budget}
\end{equation}
where $\sigma_g$ is the batch root-mean-square and $\lambda$ controls how strongly IG magnitude reshapes the budget. The $\tanh$ prevents a single extreme-gain turn from monopolizing the budget, while the additive constant guarantees that every turn, including the response, retains a baseline share of teacher supervision, so answer synthesis remains distilled rather than discarded. Setting $\lambda=0$ recovers a uniform per-turn budget $q_t=1/T$, which we use as an equal-turn ablation; our default is $\lambda=0.1$.

To implement $q_t$ without changing the trainer's existing token-level loss reduction, we rescale each sampled token's OPD reward by the ratio between its calibrated and original turn-level shares:
\begin{equation}
A^{\mathrm{ITC}}_{t,i}=\frac{N q_t}{n_t}\,D^{\mathrm{OPD}}_{t,i},
\qquad\text{so that}\qquad
\frac{1}{N}\sum_{t=1}^{T}\sum_{i\in\mathcal{I}_t}A^{\mathrm{ITC}}_{t,i}
=\sum_{t=1}^{T}q_t\Big(\frac{1}{n_t}\sum_{i\in\mathcal{I}_t}D^{\mathrm{OPD}}_{t,i}\Big).
\label{eq:itc_token}
\end{equation}
Compared to Eq.~\ref{eq:opd_decomposition}, ITC leaves the direction prior untouched and replaces only the budget prior $n_t/N$ with $q_t$. ITC deliberately uses the \textit{absolute} gain: whether helpful or harmful, a large $\vert{}g_t\vert{}$ means the turn mattered and merits more supervision, while its sign is handled separately by IQC.

\vspace{-5pt}
\subsection{Information-Guided Query-Preference Calibration}
\label{sec:iqc}
Within a search turn $\tau_t$, the direction prior ($\log\frac{\pi_T}{\pi_S}$) is computed before evidence arrives, pushing every token toward the teacher's wording. Therefore, a student query that brings decisive evidence can still be suppressed. IQC mitigates this by using the signed gain $g_t$ as a post-retrieval calibration signal. The correction applies only to action-critical tokens (e.g., tool names, arguments, and the query string), excluding formatting tokens like \texttt{<tool\_call>}. We denote this set by $\mathcal{M}_t\subseteq\mathcal{I}_t$ with $m_t=|\mathcal{M}_t|$. Since $g_t$ is a property of the whole tool call rather than individual tokens, we treat $a_t$ as one macro-action with realized utility $g_t$ and split the correction evenly over $\mathcal{M}_t$:
\begin{equation}
A^{\mathrm{IQC}}_{t,i}=\mathbf{1}\!\left[i\in\mathcal{M}_t\right]\frac{\alpha_k\,g_t}{m_t},
\qquad
\sum_{i\in\mathcal{M}_t}A^{\mathrm{IQC}}_{t,i}=\alpha_k\,g_t.
\label{eq:iqc}
\end{equation}
The second identity ensures that the total correction assigned to a search decision equals its scaled realized gain, regardless of the exact number of tokens used to express the action. This uniform allocation operates directly at the macro-action level, ensuring robust credit assignment across the entire tool invocation while preventing length-induced token biases. Accordingly, a positive $g_t$ increases the OPD signal on the decision tokens, whereas a negative $g_t$ decreases it, correcting the teacher's pre-retrieval preferences when they disagree with the utility observed after retrieval.

\textbf{Cosine-ramped evidence coefficient.} Early in training, as the student explores search space, raw posterior signals exhibit high natural variance. Rather than over-fitting to early noisy rollouts, we introduce a cosine schedule (further analyzed in Section~\ref{sec:schedule}) to smoothly transition the student from initial exploration to evidence-calibrated refinement over steps $k=0,\dots,K$: 
\begin{equation}
\alpha_k=\frac{\alpha_{\max}}{2}\Big[1-\cos\Big(\pi\,\frac{k}{K}\Big)\Big],
\label{eq:alpha_schedule}
\end{equation}

\textbf{Power-posterior interpretation.} The additive form of Eq.~\ref{eq:iqc} is not an arbitrary reward bonus. Consider an idealized sequence-level OPD objective, where the unmodified per-trajectory reward is $\log p_T(\tau\mid x)-\log p_S(\tau\mid x)$, and write $\Psi_t=\exp(\Phi_t)$. By telescoping, adding $\alpha_k\sum_t g_t$ yields:
\begin{equation}
\log p_T(\tau\mid x)-\log p_S(\tau\mid x)+\alpha_k\sum_{t}g_t
=\log\frac{p_T(\tau\mid x)\,\Psi_{T-1}(\tau)^{\alpha_k}}{p_S(\tau\mid x)}+\text{const},
\label{eq:power_posterior}
\end{equation}
i.e., the student is distilled toward a generalized power posterior $\tilde{p}_{T,k}(\tau\mid x,y^\star)\propto p_T(\tau\mid x)\,\Psi_{T-1}(\tau)^{\alpha_k}$ that tilts the teacher's prior over trajectories by a confidence pseudo-likelihood of the gold answer, with $\alpha_k$ acting as a likelihood power. IQC can therefore be read as distilling from a teacher whose preferences have been updated by the evidence the student actually retrieved. Our practical objective departs from this idealized form in three ways: the budget is calibrated by ITC, $\Psi_t$ is a length-normalized pseudo-likelihood rather than a full sequence likelihood, and the correction is assigned to $\mathcal{M}_t$ at the macro-action level. The full derivation is deferred to Appendix~\ref{app:power_posterior}.

\subsection{Overall Objective}
\label{sec:objective}
Combining the two calibrations gives the per-token advantage used to train \ours:
\begin{equation}
\boxed{\;
A_{t,i}
=\underbrace{\frac{N q_t}{n_t}\,D^{\mathrm{OPD}}_{t,i}}_{\text{ITC: IG-calibrated turn credit}}
\;+\;
\underbrace{\mathbf{1}\!\left[i\in\mathcal{M}_t\right]
\frac{\alpha_k g_t}{m_t}}_{\text{IQC: posterior evidence correction}}
\;}
\label{eq:overall}
\end{equation}
During each update, we treat these calibrated advantages as fixed supervision signals for sampled tokens. We then optimize the student using the following token-level policy-gradient surrogate:
\begin{equation}
\widehat{\mathcal{L}}_{\ours}(\theta)
=-\mathbb{E}_{x\sim\mathcal{D},\,\tau\sim\pi_{S,\theta_k}}
\left[
\frac{1}{N}\sum_t\sum_{i\in\mathcal{I}_t}
\operatorname{sg}(A_{t,i})
\log\pi_{S,\theta}(u_{t,i}\mid z_{t,i})
\right],
\label{eq:surrogate}
\end{equation}
where $\theta_k$ denotes the policy used to sample the rollout and
$\operatorname{sg}$ stops gradients through the advantage. Thus, ITC
reallocates the normalized teacher-supervision budget across turns, while
IQC reinforces or suppresses sampled query tokens according to the signed
post-retrieval gain. The IQC scores sum to $\alpha_k g_t$ within each
search turn, independently of $q_t$, and the gains telescope across turns.
Computing $g_t$ and $q_t$ requires only one additional lightweight teacher scoring pass of
the reference answer per turn but no additional rollouts. The training procedure is given in Algorithm~\ref{alg:igopd} in
Appendix~\ref{app:algorithm}.

\vspace{-5pt}
\section{Experiment}
\label{experiment}
\vspace{-5pt}

\begin{table}[t]
\centering
\small
\setlength{\tabcolsep}{4.2pt}
\renewcommand{\arraystretch}{1.0}
\caption{Performance on six multimodal search benchmarks. All benchmarks are
out-of-distribution with respect to our training data. \textbf{Bold} and
\underline{underline} mark the best and second-best score.}
\label{tab:main}
\begin{tabular}{l cccccc c}
\toprule
\multirow{2}{*}{\textbf{Method}}
& \multicolumn{6}{c}{\textbf{Multimodal Search QA}}
& \multirow{2}{*}{\textbf{Avg.}} \\
\cmidrule(lr){2-7}
& SimpleVQA & LiveVQA & MMSearch & MMSearch+
& FVQA & BrowseComp-VL & \\
\midrule

\rowcolor{grpA}
\multicolumn{8}{l}{\textit{Direct Reasoning}} \\
GPT-5
& 61.6 & 44.4 & 35.1 & 19.1 & 54.4 & 48.6 & 43.9 \\
Gemini-2.5-Pro
& 63.0 & 60.3 & 39.8 & 14.5 & 60.7 & 43.1 & 46.9 \\
Claude-4-Sonnet
& 50.9 & 38.5 & 18.7 & 4.0 & 35.3 & 29.3 & 29.5 \\

\midrule
\rowcolor{grpB}
\multicolumn{8}{l}{\textit{Agentic Workflow --- OpenSearch-VL-8B (Teacher)}} \\
GRPO Enhanced
& 71.6 & 66.0 & 64.3 & 19.5 & 70.7 & 44.7 & 56.1 \\

\midrule
\rowcolor{grpC}
\multicolumn{8}{l}{\textit{Agentic Workflow --- Qwen3-VL-2B-Instruct (Student)}} \\
Zero-Shot
& 21.0 & 8.0 & 14.7 & 1.8 & 8.0 & 1.3 & 9.1 \\
SFT
& 57.0 & 52.7 & 51.0 & \underline{12.2}
& 57.0 & 29.0 & 43.2 \\
GRPO
& 61.7 & 52.0 & 53.7 & 10.8
& 60.3 & 31.7 & 45.0 \\
IGPO
& 60.3 & \underline{53.7} & 55.3 & \underline{12.2}
& 61.0 & \underline{33.3} & 46.0 \\
OPD
& 62.3 & 52.0 & 55.7 & 11.0
& \underline{62.0} & 33.0 & 46.0 \\
TCOD
& \textbf{63.0} & 53.3 & \underline{57.7} & 10.4
& 61.7 & 31.3 & \underline{46.2} \\
\rowcolor{ourrow}
\textbf{IG-OPD (Ours)}
& \underline{62.7} & \textbf{56.3} & \textbf{62.0} & \textbf{14.4}
& \textbf{63.7} & \textbf{37.3} & \textbf{49.4} \\

\midrule
\rowcolor{grpC}
\multicolumn{8}{l}{\textit{Agentic Workflow --- Qwen3-VL-4B-Instruct (Student)}} \\
Zero-Shot
& 58.0 & 51.3 & 49.3 & 9.0
& 57.0 & 24.0 & 41.4 \\
SFT
& 65.0 & 61.3 & 54.0 & 16.7
& 64.7 & 38.0 & 50.0 \\
GRPO
& \underline{66.7} & 62.3 & 60.3 & 15.3
& 65.0 & 41.7 & 51.9 \\
IGPO
& 66.0 & \underline{63.7} & 61.0 & \underline{17.2}
& 66.7 & 43.3 & 53.0 \\
OPD
& \underline{66.7} & \underline{63.7} & \underline{62.0} & 16.2
& \textbf{67.7} & \underline{43.7} & \underline{53.3} \\
TCOD
& 63.7 & 63.3 & 59.7 & 16.2
& \underline{67.3} & 42.7 & 52.2 \\
\rowcolor{ourrow}
\textbf{IG-OPD (Ours)}
& \textbf{67.7} & \textbf{65.0} & \textbf{66.7} & \textbf{17.6}
& \underline{67.3} & \textbf{45.7} & \textbf{55.0} \\

\bottomrule
\end{tabular}
\end{table}

\subsection{Experiment Setup}

\textbf{Benchmarks \& Metrics.}
We evaluate on six multimodal web-search benchmarks spanning single-hop factual lookup, multi-hop retrieval, and long-horizon deep search: SimpleVQA~\citep{cheng2025simplevqa}, LiveVQA~\citep{fu2026seeking}, MMSearch~\citep{jiang2024mmsearch}, MMSearch+~\citep{tao2026mmsearch}, FVQA test split~\citep{wang2017fvqa}, and BrowseComp-VL~\citep{geng2026webwatcher}. Following \citet{chen2026opensearch}, we report accuracy judged against reference answers by GPT-4o.

\textbf{Baselines.} To evaluate the effectiveness of our proposed \ours, we utilize several methods spanning from RL to OPD for comparison, including vanilla decoding, \textit{SFT}, \textit{GRPO}~\citep{shao2024deepseekmath}, \textit{IGPO}~\citep{wang2026information}, \textit{OPD}~\citep{agarwal2024policy}, and \textit{TCOD}~\citep{wang2026tcod}.

\textbf{Implementation Details.}
We adopt OpenSearch-VL-8B~\citep{chen2026opensearch} as the teacher agent. To optimize efficiency and supervision quality, we optimize its tool set and further fine-tune it with GRPO to elevate the distillation ceiling (Appendix~\ref{app:tools}). Our students are Qwen3-VL-2B/4B-Instruct~\citep{bai2025qwen3}, which share the teacher's vocabulary. Before distillation, we cold-start the student agent via SFT on 10K trajectories sampled from SearchVL-SFT-36k to equip the student with basic agentic search capability. We then perform OPD training using 4K difficulty-stratified samples sourced from Search-VL-RL-8K~\citep{chen2026opensearch}, VDR-Bench-2K~\citep{zeng2026vision}, and FVQA train split~\citep{wang2017fvqa}. Notably, both training and evaluation are conducted in a live web environment, with search actions served by the Serper and Jina APIs. All training runs are executed on $8\times\text{NVIDIA H100 GPUs}$. Unless stated otherwise, \ours uses $\lambda=0.1$, $\alpha_{\max}=10$, and a $K=60$-step cosine schedule. Appendix~\ref{app:training} provides full training and implementation details.

% =====================================================================
\vspace{-5pt}
\subsection{Can \ours Transfer Deep-Search Capability to Compact Students?}
\label{sec:main_results}
\vspace{-5pt}

Table~\ref{tab:main} reports results on all six benchmarks for both student scales.

\textbf{\ours achieves the best average at both scales and lets a 4B student rival its 8B teacher.}
Under the same cold-start model and training sources, \ours improves average accuracy over vanilla OPD by 3.4 points at 2B (49.4 vs.\ 46.0) and by 1.7 points at 4B (55.0 vs.\ 53.3), ranking first on five of the six benchmarks in both student groups. Notably, the 4B student surpasses the 8B teacher on MMSearch (66.7 vs.\ 64.3) and BrowseComp-VL (45.7 vs.\ 44.7), effectively matching the teacher's capability overall with half the scale. Although the teacher provides dense supervision, IQC calibrates query-level signals using evidence actually retrieved by the student rather than relying solely on the teacher's prior preference for query wording. The student effectively breaks the OPD ceiling, allowing the student to outperform its teachers. Both \ours students also outperform all direct-reasoning frontier models (e.g., 49.4 for our 2B student vs.\ 46.9 for Gemini-2.5-Pro), confirming that these benchmarks genuinely require active retrieval rather than parametric recall.

\textbf{Gains concentrate on search-intensive benchmarks where EBS is most consequential.}
Compared with OPD, \ours achieves the largest gains on search-intensive benchmarks: MMSearch (+6.3 at 2B / +4.7 at 4B), BrowseComp-VL (+4.3 / +2.0), LiveVQA (+4.3 / +1.3), and MMSearch+ (+3.4 / +1.4). In contrast, on shallow-search benchmarks with limited retrieval steps, \ours maintains consistently competitive performance (e.g., SimpleVQA and FVQA). This directly confirms our intuition: with fewer search turns, there is minimal supervision budget to misallocate (ITC) and few query directions to correct (IQC). As trajectories involve more search turns, vanilla OPD accumulates severe supervision misalignment (Figure~\ref{fig:ebs_diagnosis}), making our calibration increasingly crucial.

\textbf{Calibrating supervision helps where a turn curriculum and outcome rewards do not.}
TCOD's turn scheduling offers marginal gains at 2B (46.2 vs.\ 46.0) and underperforms OPD at 4B (52.2 vs.\ 53.3), suggesting that delaying supervision alone is insufficient without calibrating credit magnitude or direction. Outcome-based GRPO also trails OPD at both scales (45.0/51.9), consistent with the difficulty of assigning credit through sparse trajectory-level rewards. Crucially, IGPO (incorporating IG into policy optimization) reaches only 46.0 at 2B and 53.0 at 4B, trailing IG-OPD by 3.4 and 2.0 points. Although its turn-level feedback improves over outcome-based GRPO, it only matches OPD at 2B and falls slightly below it at 4B. These results suggest that the benefit of information gain depends on how it enters the learning signal: using it to calibrate dense teacher supervision preserves token-level guidance while adjusting supervision according to retrieved evidence. In our setting, this combination of ITC and IQC is more effective than using IG as a turn-level policy reward.

% =====================================================================
\vspace{-5pt}
\subsection{Which Calibration Matters?}
\label{sec:ablation}
\vspace{-5pt}

% Table (left) and one two-panel figure (right) share a row. The outer float only
% controls placement; \captionof keeps a Table number on the left and a Figure
% number on the right. Both blocks are vertically centred ([c]) so the caption
% above the table and the caption below the figure do not fight each other.
\begin{figure}[t]
\centering
\begin{minipage}[c]{0.44\textwidth}
\centering
\captionof{table}{Component ablation on MMSearch (2B student).
The IQC-only variant isolates query calibration, while the remaining
rows progressively build \ours from OPD.}
\label{tab:ablation}
\small
\setlength{\tabcolsep}{5pt}
\renewcommand{\arraystretch}{1.08}
\begin{tabular}{@{}l r r@{}}
\toprule
\textbf{Variant} & \textbf{Acc.} & $\boldsymbol{\Delta}$ \\
\midrule
OPD (budget $n_t/N$)
    & 55.7 & -- \\
$+$ IQC only (no ITC)
    & 56.7 & $+1.0$ \\
\addlinespace[2pt]
\cmidrule(lr){1-3}
\addlinespace[1pt]
$+$ turn normalization ($\lambda{=}0$)
    & 58.3 & $+2.6$ \\
$+$ ITC with signed $g_t$
    & 59.3 & $+3.6$ \\
$+$ ITC with $|g_t|$ (default)
    & 60.0 & $+4.3$ \\
\rowcolor{ourrow}
$+$ IQC ($=$ \ours)
    & \textbf{62.0} & $\boldsymbol{+6.3}$ \\
\bottomrule
\end{tabular}
\end{minipage}\hfill
\begin{minipage}[c]{0.54\textwidth}
\centering
\includegraphics[width=\linewidth]{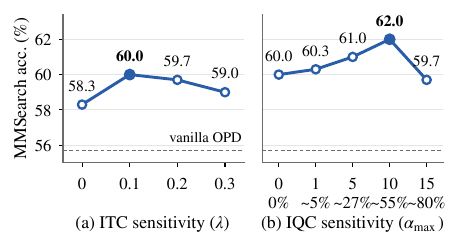}
\vspace{-15pt}
\captionof{figure}{Sensitivity on MMSearch (2B student).
(a) ITC strength $\lambda$ with IQC disabled.
(b) IQC strength $\alpha_{\max}$ with ITC at its default, with secondary x-ticks indicating relative correction magnitude to OPD.}
\vspace{-10pt}
\label{fig:sensitivity}
\end{minipage}
\end{figure}

\textbf{Both priors are harmful, and their corrections are complementary.}
Table~\ref{tab:ablation} isolates and progressively combines the components of \ours. Replacing the length-induced budget $n_t/N$ with a uniform per-turn budget improves OPD by 2.6 points (58.3), surpassing TCOD (57.7); the token-volume prior is harmful even before information gain is introduced. Modulating this budget by the realized information magnitude adds 1.7 points (60.0), with absolute gain outperforming signed gain (59.3). This supports using magnitude to determine \textit{how much} a turn is supervised, while IQC handles direction. IQC alone yields a 1.0-point gain (56.7), whereas adding it after ITC contributes 2.0 points (62.0). The two calibrations thus address distinct but coupled failure modes: ITC reallocates supervision toward informative turns but cannot correct a teacher-disfavored query, while IQC corrects its direction but is less effective when the original budget under-credits search turns.

\textbf{\ours is robust to $\lambda$ and benefits from a moderate evidence coefficient.}
Figure~\ref{fig:sensitivity}(a) shows that any $\lambda\in[0.1,0.3]$ outperforms the uniform budget, with the best result at $\lambda=0.1$ and a gentle decline thereafter (59.7, 59.0); since $\tanh(\cdot)\le 1$, even $\lambda=0.3$ bounds the ratio between the most and least credited turns by $1.3$, so ITC reshapes rather than overturns the teacher's supervision. Figure~\ref{fig:sensitivity}(b) shows that accuracy increases monotonically with $\alpha_{\max}$ up to $10$ (60.0$\rightarrow$62.0), at which point the IQC correction amounts to roughly half of the OPD signal, and drops to 59.7 at $\alpha_{\max}=15$ as excessively large corrections disrupt the underlying dense teacher prior, confirming that posterior calibration acts best as a complementary guidance rather than a complete replacement. These observations suggest that the posterior evidence should remain a moderate correction. 

% =====================================================================
\vspace{-7pt}
\subsection{Does \ours Resolve Evidence-Blind Supervision?}
\label{sec:ism_analysis}

We now return to the two symptoms in Figure~\ref{fig:ebs_diagnosis} and measure them again after calibration.

\begin{figure}[t]
\centering
\vspace{-10pt}
\includegraphics[width=\textwidth]{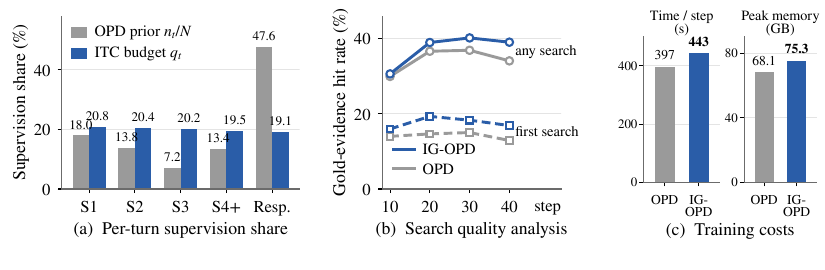}
\vspace{-25pt}
\caption{\ours effectively mitigates evidence-blind supervision. (a) Per-turn supervision share under OPD prior $n_t/N$ and ITC budget $q_t$. (b) Gold-evidence retrieval rate across training (first and any search turns). (c) Training time per step and peak GPU memory for OPD and \ours.}
\vspace{-10pt}
\label{fig:ebs_after}
\end{figure}

\textbf{ITC shifts supervision budget from token volume to information criticality.}
Figure~\ref{fig:ebs_after}(a) contrasts the length-induced share $n_t/N$ with the ITC budget $q_t$. Under the prior, the response turn absorbs 47.6\% of supervision while the third search turn receives 7.2\%; under ITC, the four search groups receive 19.5\%--20.8\% each (80.9\% total vs.\ 52.4\%), ordered by their realized information magnitude (Search 1 highest). The response turn retains 19.1\% (near uniform share $1/T$), ensuring answer synthesis is distilled rather than discarded via the additive constant in Eq.~\ref{eq:itc_budget}.

\textbf{IQC improves search quality and precision.}
Figure~\ref{fig:ebs_after}(b) tracks how often a rollout's searches retrieve the gold evidence during training. \ours raises the hit rate of the \emph{first} search from 12.9\% to 16.9\% and of \emph{any} search from 34.1\% to 39.1\% by step 40, with the gap widening as $\alpha_k$ ramps up, while the number of searches per rollout stays essentially unchanged (4.35 vs.\ 4.26). Calibrated supervision therefore improves the queries the student issues rather than their number, consistent with the 4B student's advantage over its teacher on the multi-hop benchmarks in Sec.~\ref{sec:main_results}.

\textbf{Evidence calibration adds modest training overhead.}
Figure~\ref{fig:ebs_after}(c) compares training costs of OPD and \ours on the same hardware. Time per step rises from 397 to 443 seconds (+11.6\%), while peak GPU memory increases from 68.1 to 75.3 GB (+10.6\%). This cost arises during training, where post-retrieval information gain is evaluated to calibrate supervision. The increase is moderate relative to the performance gains reported above. At inference, the student follows the same search workflow without computing information gain, so \ours adds no inference overhead.
% =====================================================================
\vspace{-7pt}
\subsection{Is Post-Retrieval IG Reliable, and Why Ramp It?}
\label{sec:schedule}

\begin{wrapfigure}[15]{r}{0.38\textwidth}
\vspace{-0.8\baselineskip}
\centering
\captionsetup{font=small,skip=4pt}
\captionof{table}{Results of IG reliability.}
\label{tab:ig_reliability}
\small
\setlength{\tabcolsep}{4pt}
\renewcommand{\arraystretch}{1.00}
\begin{tabular}{@{}lcc@{}}
\toprule
\textbf{Search position}
& \shortstack{\textbf{IG$>0$}\\$\rightarrow$ correct}
& \shortstack{\textbf{IG$<0$}\\$\rightarrow$ wrong} \\
\midrule
First search
& 87.5\% & 76.2\% \\
Later searches
& 54.3\% & 85.7\% \\
\bottomrule
\end{tabular}
\vspace{8pt}
\includegraphics[width=0.95\linewidth]{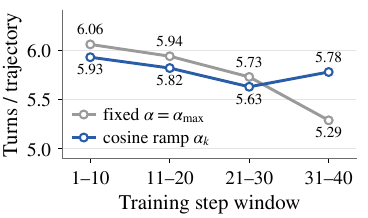}
\vspace{-10pt}
\captionof{figure}{Results of search suppression.}
\label{fig:schedule}
\vspace{-0.5\baselineskip}
\end{wrapfigure}

\noindent \textbf{IG reliably reflects the final outcome, but its predictive direction
depends on search position.}
Since IG drives both calibrations, Table~\ref{tab:ig_reliability} assesses its alignment with final-answer correctness. Agreement reaches 82.2\% on the first search and 70.0\% on later ones. The errors exhibit clear position asymmetry: positive IG on the first search strongly predicts success (87.5\%), whereas 
 negative IG  on later searches correctly predicts failure (85.7\%). Focusing on the top 30\% largest gains raises both rates to $\sim$90\%. Early on, a positive jump from low potential $\Phi_t$ is decisive; later, small fluctuations are mostly verification noise, while large drops signal misleading evidence.

\textbf{The cosine ramp prevents early search suppression.}
Early in distillation, weak student retrievals yield predominantly negative IG. Unramped IQC ($\alpha_{\max}$ at step 0) turns this into a general search penalty, dropping average rollout length from 6.06 to 5.29 turns over the first 40 steps on training set (Figure~\ref{fig:schedule}). The cosine ramp keeps IQC weak initially while the student acquires teacher search behavior, ensuring IG refines rather than suppresses search once fully active (preserving 5.78 turns). The ramp thus acts as a curriculum from imitation to evidence-calibrated distillation.

\vspace{-10pt}
\section{Conclusion}
\label{conclusion}
\vspace{-8pt}

We have presented \ours, a novel distillation framework designed to address Evidence-Blind Supervision (EBS) in deep search agents. By diagnosing the structural mismatches in traditional On-Policy Distillation (OPD), \ours introduces two synergistic, evidence-aware calibrations: Information-guided Turn-credit Calibration (ITC), which reallocates supervision based on localized information gain rather than token length, and Information-guided Query-preference Calibration (IQC), which aligns query update directions with the realized utility of retrieved evidence. Extensive evaluation across six OOD multimodal search benchmarks demonstrates that \ours significantly outperforms existing distillation and reinforcement learning baselines. Notably, our 4B student effectively bridges the capacity gap, rivaling and even surpassing its 8B teacher without incurring inference-time overhead. Moving forward, extending post-retrieval evidence calibration to other environments represents a highly promising direction for further scaling autonomous web agents.

\section*{AI Use Statement}

We used generative AI tools to support literature discovery, code development
and debugging, exploratory analysis of experimental results, figure and table
preparation, and language editing. The authors independently verified the
relevant literature, implementations, numerical results, and manuscript
content. All methodological choices, scientific claims, and conclusions were
made by the authors, who take full responsibility for the final manuscript and
its accompanying artifacts.

\section*{Reproducibility Statement}

We provide the detailed derivation and complete
IG-OPD algorithm in Appendix~\ref{app:method}. The agent environment, teacher
preparation, training configurations, and data construction procedures are
documented in Appendix~\ref{app:experimental_setup}, while the rollout and
evaluation prompts are reproduced in Appendix~\ref{app:prompts}. The benchmark setup,
evaluation metrics, baselines, and sampling protocol are described in
Section~\ref{experiment}.

\bibliography{iclr2027_conference}

@article{wu2025mmsearch,
  title={Mmsearch-r1: Incentivizing lmms to search},
  author={Wu, Jinming and Deng, Zihao and Li, Wei and Liu, Yiding and You, Bo and Li, Bo and Ma, Zejun and Liu, Ziwei},
  journal={arXiv preprint arXiv:2506.20670},
  year={2025}
}

@article{chen2026opensearch,
  title={Opensearch-vl: An open recipe for frontier multimodal search agents},
  author={Chen, Shuang and Feng, Kaituo and Chen, Hangting and Huang, Wenxuan and Dai, Dasen and Shou, Quanxin and Lin, Yunlong and Yue, Xiangyu and Gao, Shenghua and Pang, Tianyu},
  journal={arXiv preprint arXiv:2605.05185},
  year={2026}
}

@article{li2026hypereyes,
  title={HyperEyes: Dual-Grained Efficiency-Aware Reinforcement Learning for Parallel Multimodal Search Agents},
  author={Li, Guankai and Chen, Jiabin and Xu, Yi and Zhang, Xichen and Lu, Yuan},
  journal={arXiv preprint arXiv:2605.07177},
  year={2026}
}

@article{lewis2020retrieval,
  title={Retrieval-augmented generation for knowledge-intensive nlp tasks},
  author={Lewis, Patrick and Perez, Ethan and Piktus, Aleksandra and Petroni, Fabio and Karpukhin, Vladimir and Goyal, Naman and K{\"u}ttler, Heinrich and Lewis, Mike and Yih, Wen-tau and Rockt{\"a}schel, Tim and others},
  journal={Advances in neural information processing systems},
  volume={33},
  pages={9459--9474},
  year={2020}
}

@inproceedings{mallen2023not,
  title={When not to trust language models: Investigating effectiveness of parametric and non-parametric memories},
  author={Mallen, Alex and Asai, Akari and Zhong, Victor and Das, Rajarshi and Khashabi, Daniel and Hajishirzi, Hannaneh},
  booktitle={Proceedings of the 61st annual meeting of the association for computational linguistics (volume 1: Long papers)},
  pages={9802--9822},
  year={2023}
}

@article{jiang2024mmsearch,
  title={Mmsearch: Benchmarking the potential of large models as multi-modal search engines},
  author={Jiang, Dongzhi and Zhang, Renrui and Guo, Ziyu and Wu, Yanmin and Lei, Jiayi and Qiu, Pengshuo and Lu, Pan and Chen, Zehui and Fu, Chaoyou and Song, Guanglu and others},
  journal={arXiv preprint arXiv:2409.12959},
  year={2024}
}

@inproceedings{agarwal2024policy,
  title={On-policy distillation of language models: Learning from self-generated mistakes},
  author={Agarwal, Rishabh and Vieillard, Nino and Zhou, Yongchao and Stanczyk, Piotr and Ramos Garea, Sabela and Geist, Matthieu and Bachem, Olivier},
  booktitle={International Conference on Learning Representations},
  volume={2024},
  pages={21246--21263},
  year={2024}
}

@article{ko2026scaling,
  title={Scaling reasoning efficiently via relaxed on-policy distillation},
  author={Ko, Jongwoo and Abdali, Sara and Kim, Young Jin and Chen, Tianyi and Cameron, Pashmina},
  journal={arXiv preprint arXiv:2603.11137},
  year={2026}
}

@article{hao2026stabilizing,
  title={Stabilizing On-Policy Distillation for MLLM Reasoning with Global Normalization},
  author={Hao, Dongze and Jin, Zhiwei and Chen, Chen and Lu, Haonan},
  journal={arXiv preprint arXiv:2606.09091},
  year={2026}
}

@article{xie2026trust,
  title={Trust Region Policy Distillation},
  author={Xie, Zhengpeng and Zhang, Li Lyna and Xie, Zeke and Yang, Mao},
  journal={arXiv preprint arXiv:2607.04751},
  year={2026}
}

@article{jin2026entropy,
  title={Entropy-aware on-policy distillation of language models},
  author={Jin, Woogyeol and Min, Taywon and Yang, Yongjin and Kadhe, Swanand Ravindra and Zhou, Yi and Wei, Dennis and Baracaldo, Nathalie and Lee, Kimin},
  journal={arXiv preprint arXiv:2603.07079},
  year={2026}
}

@article{tan2026atod,
  title={ATOD: Annealed Turn-aware On-policy Distillation for Multi-turn Autonomous Agents},
  author={Tan, Qitai and Zong, Zefang and Li, Yang and Chen, Peng},
  journal={arXiv preprint arXiv:2606.27814},
  year={2026}
}

@article{xu2025kdrl,
  title={Kdrl: Post-training reasoning llms via unified knowledge distillation and reinforcement learning},
  author={Xu, Hongling and Zhu, Qi and Deng, Heyuan and Li, Jinpeng and Hou, Lu and Wang, Yasheng and Shang, Lifeng and Xu, Ruifeng and Mi, Fei},
  journal={arXiv preprint arXiv:2506.02208},
  year={2025}
}

@article{wang2026tcod,
  title={Tcod: Exploring temporal curriculum in on-policy distillation for multi-turn autonomous agents},
  author={Wang, Jiaqi and Zhang, Wenhao and Shi, Weijie and Li, Yaliang and Cheng, James},
  journal={arXiv preprint arXiv:2604.24005},
  year={2026}
}

@article{zhang2025deep,
  title={Deep research: A survey of autonomous research agents},
  author={Zhang, Wenlin and Li, Xiaopeng and Zhang, Yingyi and Jia, Pengyue and Wang, Yichao and Guo, Huifeng and Liu, Yong and Zhao, Xiangyu},
  journal={arXiv preprint arXiv:2508.12752},
  year={2025}
}

@article{lin2025comprehensive,
  title={A comprehensive survey on reinforcement learning-based agentic search: Foundations, roles, optimizations, evaluations, and applications},
  author={Lin, Minhua and Wu, Zongyu and Xu, Zhichao and Liu, Hui and Tang, Xianfeng and He, Qi and Aggarwal, Charu and Zhang, Xiang and Wang, Suhang},
  journal={arXiv preprint arXiv:2510.16724},
  year={2025}
}

@inproceedings{geng2026webwatcher,
 author = {Geng, Xinyu and Xia, Peng and Zhang, Zhen and Wang, Xinyu and Wang, Qiuchen and Ding, Ruixue and Wang, Chenxi and Wu, Jialong and Li, Kuan and Zhao, Yida and Yin, Huifeng and Jiang, Yong and Xie, Pengjun and Huang, Fei and Yao, Huaxiu and Fung, Yi R. and Zhou, Jingren},
 booktitle = {International Conference on Learning Representations},
 editor = {C. Vondrick and B. Hariharan and C. Raffel and L. Pinto and D. Yang and A. Faust},
 pages = {61240--61271},
 title = {WebWatcher: Breaking New Frontiers of Vision-Language Deep Research Agent},
 volume = {2026},
 year = {2026}
}

@article{hong2025deepeyesv2,
  title={Deepeyesv2: Toward agentic multimodal model},
  author={Hong, Jack and Zhao, Chenxiao and Zhu, ChengLin and Lu, Weiheng and Xu, Guohai and Yu, Xing},
  journal={arXiv preprint arXiv:2511.05271},
  year={2025}
}

@article{xu2026tip,
  title={Tip: Token importance in on-policy distillation},
  author={Xu, Yuanda and Sang, Hejian and Zhou, Zhengze and He, Ran and Wang, Zhipeng and Geramifard, Alborz},
  journal={arXiv preprint arXiv:2604.14084},
  year={2026}
}

@article{zheng2026scope,
  title={Scope: Signal-calibrated on-policy distillation enhancement with dual-path adaptive weighting},
  author={Zheng, Binbin and Ma, Xing and Liang, Yiheng and Ruan, Jingqing and Fu, Xiaoliang and Lin, Kepeng and Zhu, Benchang and Zeng, Ke and Cai, Xunliang},
  journal={arXiv preprint arXiv:2604.10688},
  year={2026}
}

@article{huang2025selectkd,
  title={SelecTKD: Selective Token-Weighted Knowledge Distillation for LLMs},
  author={Huang, Haiduo and Song, Jiangcheng and Zhang, Yadong and Ren, Pengju},
  journal={arXiv preprint arXiv:2510.24021},
  year={2025}
}

@article{li2026filter,
  title={Filter, then reweight: Rethinking optimization granularity in on-policy distillation},
  author={Li, Yuying and Zheng, Leqi and Yu, Yongzi and Zhou, Wenrui and Zhong, Xuchang and Hu, Xing and Jin, Jing and Yuan, Hangjie and Feng, Tao},
  journal={arXiv preprint arXiv:2606.02684},
  year={2026}
}

@inproceedings{xu2026rlkd,
  title={Rlkd: Distilling llms’ reasoning via reinforcement learning},
  author={Xu, Shicheng and Pang, Liang and Zhu, Yunchang and Gu, Jia and Wei, Zihao and Deng, Jingcheng and Pan, Feiyang and Shen, Huawei and Cheng, Xueqi},
  booktitle={Proceedings of the AAAI Conference on Artificial Intelligence},
  volume={40},
  pages={34151--34159},
  year={2026}
}

@article{zhang2026reinforcement,
  title={Reinforcement-aware knowledge distillation for llm reasoning},
  author={Zhang, Zhaoyang and Jiang, Shuli and Shen, Yantao and Zhang, Yuting and Ram, Dhananjay and Yang, Shuo and Tu, Zhuowen and Xia, Wei and Soatto, Stefano},
  journal={arXiv preprint arXiv:2602.22495},
  year={2026}
}

@article{zhong2026sod,
  title={Sod: Step-wise on-policy distillation for small language model agents},
  author={Zhong, Qiyong and Zheng, Mao and Song, Mingyang and Lin, Xin and Sun, Jie and Jiang, Houcheng and Wang, Xiang and Fang, Junfeng},
  journal={arXiv preprint arXiv:2605.07725},
  year={2026}
}

@article{li2026policy,
  title={On-Policy Distillation with Curriculum Turn-level Guidance for Multi-turn Agents},
  author={Li, Gengsheng and Zheng, Mao and Song, Mingyang and Liu, Ruiqi and Yang, Tianyu and Sun, Jie and Zhong, Qiyong and Guo, Haiyun and Fang, Junfeng and Zhang, Dan and others},
  journal={arXiv preprint arXiv:2606.15912},
  year={2026}
}

@article{zhou2026turnopd,
  title={TurnOPD: Making On-Policy Distillation Turn-Aware for Efficient Long-Horizon Agent Training},
  author={Zhou, Yuhang and Zheng, Kai and Li, Haoling and Peng, Dengyun and Xu, Can and Chen, Jingjing},
  journal={arXiv preprint arXiv:2607.05804},
  year={2026}
}

@article{zhou2026sage,
  title={SAGE-OPD: Selective Agent-Guided Intervention for Multi-Turn On-Policy Distillation},
  author={Zhou, Yuhang and Zhang, Lizhu and Wu, Yifan and Wang, Mingyi and Peng, Bo and Liu, Jiayi and Fan, Xiangjun and Zhao, Zhuokai},
  journal={arXiv preprint arXiv:2606.19659},
  year={2026}
}

@article{liao2026multi,
  title={Multi-Turn On-Policy Distillation with Prefix Replay},
  author={Liao, Baohao and Dong, Hanze and Monz, Christof and Xu, Xinxing and Dong, Li and Wei, Furu},
  journal={arXiv preprint arXiv:2607.04763},
  year={2026}
}

@inproceedings{cheng2025simplevqa,
  title={Simplevqa: Multimodal factuality evaluation for multimodal large language models},
  author={Cheng, Xianfu and Zhang, Wei and Zhang, Shiwei and Yang, Jian and Guan, Xiangyuan and Wu, Xianjie and Li, Xiang and Zhang, Ge and Liu, Jiaheng and Mai, Yuying and others},
  booktitle={2025 IEEE/CVF International Conference on Computer Vision (ICCV)},
  pages={4637--4646},
  year={2025},
  organization={IEEE}
}

@inproceedings{fu2026seeking,
 author = {Fu, Mingyang and Peng, Yuyang and Chen, Dongping and Zhou, Zetong and Liu, Benlin and Wan, Yao and Zhao, Zhou and Yu, Philip S and Krishna, Ranjay},
 booktitle = {Advances in Neural Information Processing Systems},
 doi = {10.52202/085713-2513},
 editor = {D. Belgrave and C. Zhang and H. Lin and R. Pascanu and P. Koniusz and M. Ghassemi and N. Chen},
 pages = {},
 publisher = {Curran Associates, Inc.},
 title = {Seeking and Updating with Live Visual Knowledge},
 volume = {38, Main Conference},
 year = {2025}
}

@inproceedings{tao2026mmsearch,
 author = {Tao, Xijia and Yihua, Teng and Su, Xinxing and Fu, Xinyu and Wu, Jihao and Tao, Chaofan and Liu, Ziru and Bai, Haoli and Liu, Rui and Kong, Lingpeng},
 booktitle = {International Conference on Learning Representations},
 editor = {C. Vondrick and B. Hariharan and C. Raffel and L. Pinto and D. Yang and A. Faust},
 pages = {70385--70422},
 title = {MMSearch-Plus: Benchmarking Provenance-Aware Search for Multimodal Browsing Agents},
 volume = {2026},
 year = {2026}
}

@article{wang2017fvqa,
  title={Fvqa: Fact-based visual question answering},
  author={Wang, Peng and Wu, Qi and Shen, Chunhua and Dick, Anthony and Van Den Hengel, Anton},
  journal={IEEE transactions on pattern analysis and machine intelligence},
  volume={40},
  number={10},
  pages={2413--2427},
  year={2017},
  publisher={IEEE}
}

@inproceedings{wang2026information,
  title={Information gain-based policy optimization: A simple and effective approach for multi-turn search agents},
  author={Wang, Guoqing and Dai, Sunhao and Ye, Guangze and Gan, Zeyu and Yao, Wei and Deng, Yong and Wu, Xiaofeng and others},
  booktitle={International Conference on Learning Representations},
  volume={2026},
  pages={104502--104529},
  year={2026}
}

@inproceedings{ng1999policy,
  title={Policy invariance under reward transformations: Theory and application to reward shaping},
  author={Ng, Andrew Y. and Harada, Daishi and Russell, Stuart},
  booktitle={International Conference on Machine Learning},
  pages={278--287},
  year={1999}
}

@article{shao2024deepseekmath,
  title={Deepseekmath: Pushing the limits of mathematical reasoning in open language models},
  author={Shao, Zhihong and Wang, Peiyi and Zhu, Qihao and Xu, Runxin and Song, Junxiao and Bi, Xiao and Zhang, Haowei and Zhang, Mingchuan and Li, YK and Wu, Yang and others},
  journal={arXiv preprint arXiv:2402.03300},
  year={2024}
}

@article{bai2025qwen3,
  title={Qwen3-vl technical report},
  author={Bai, Shuai and Cai, Yuxuan and Chen, Ruizhe and Chen, Keqin and Chen, Xionghui and Cheng, Zesen and Deng, Lianghao and Ding, Wei and Gao, Chang and Ge, Chunjiang and others},
  journal={arXiv preprint arXiv:2511.21631},
  year={2025}
}

@article{zeng2026vision,
  title={Vision-deepresearch benchmark: Rethinking visual and textual search for multimodal large language models},
  author={Zeng, Yu and Huang, Wenxuan and Fang, Zhen and Chen, Shuang and Shen, Yufan and Cai, Yishuo and Wang, Xiaoman and Yin, Zhenfei and Chen, Lin and Chen, Zehui and others},
  journal={arXiv preprint arXiv:2602.02185},
  year={2026}
}
\bibliographystyle{iclr2027_conference}
\clearpage

\appendix
\startcontents[appendix]

\begingroup
\hypersetup{linkcolor=appendixpurple}

\titlecontents{section}
  [2.0em]
  {\addvspace{0.8em}\color{appendixpurple}\bfseries}
  {\contentslabel{2.0em}}
  {}
  {\titlerule*[0.55pc]{.}\contentspage}

\titlecontents{subsection}
  [4.5em]
  {\color{appendixpurple}}
  {\contentslabel{3.0em}}
  {}
  {\titlerule*[0.55pc]{.}\contentspage}

\section*{\scshape Appendix Contents}
\vspace{0.4em}

\printcontents[appendix]{}{1}[2]{}
\endgroup

\clearpage

\section{Method Details and Derivations}
\label{app:method}

\subsection{Turn-wise Decomposition of OPD}
\label{app:opd_decomposition}

\paragraph{Interaction and token notation.}
For an input $x$, the student produces a trajectory
$\tau=(a_1,o_1,\ldots,a_{T-1},o_{T-1},a_T)$, where $a_t$ is a
policy-generated search action for $t<T$, $o_t$ is the corresponding tool
observation, and $a_T$ is the final response.  We write $H_{t-1}$ for the
history available immediately before $a_t$ and $H_t=(H_{t-1},a_t,o_t)$ for
the history after the observation is returned.  Each action is represented
autoregressively by generated tokens.  Let $\mathcal I_t$ denote their
positions, $n_t=|\mathcal I_t|$, and $N=\sum_{t=1}^{T}n_t$.  For
$i\in\mathcal I_t$, $u_{t,i}$ is the sampled token and $z_{t,i}$ is its
complete prefix.  Tool observations enter later prefixes as context, but are
not generated by the policy and therefore do not receive token loss.

\paragraph{From prefix matching to sampled-token OPD.}
OPD matches the student to a frozen teacher on prefixes visited by the
student itself.  At a fixed prefix $z$, the reverse-KL objective is
\begin{equation}
\mathcal K(z)
=D_{\mathrm{KL}}\!\left(\pi_S(\cdot\mid z)\,\|\,\pi_T(\cdot\mid z)\right)
=\sum_{u\in\mathcal V}\pi_S(u\mid z)
\log\frac{\pi_S(u\mid z)}{\pi_T(u\mid z)},
\label{eq:app_prefix_kl}
\end{equation}
where $\mathcal V$ is the vocabulary.  Define the sampled teacher--student
log-probability gap
\begin{equation}
D^{\mathrm{OPD}}(u,z)
=\log\pi_T(u\mid z)-\log\pi_S(u\mid z).
\label{eq:app_token_gap}
\end{equation}
For $u\sim\pi_S(\cdot\mid z)$, this is an unbiased one-sample estimator of
the negative prefix divergence:
\begin{equation}
\mathbb E_{u\sim\pi_S(\cdot\mid z)}
\big[D^{\mathrm{OPD}}(u,z)\big]
=-\mathcal K(z).
\label{eq:app_gap_expectation}
\end{equation}
Sampling therefore replaces the vocabulary-wide sum with the token already
generated during the on-policy rollout.  Averaging these samples over all
generated positions gives
\begin{equation}
\widehat{\mathcal L}_{\mathrm{OPD}}
=-\mathbb E_{x,\,\tau\sim\pi_S(\cdot\mid x)}
\left[\frac{1}{N}\sum_{t=1}^{T}\sum_{i\in\mathcal I_t}
\underbrace{\log\frac{\pi_T(u_{t,i}\mid z_{t,i})}
{\pi_S(u_{t,i}\mid z_{t,i})}}_{D^{\mathrm{OPD}}_{t,i}}\right].
\label{eq:app_sampled_opd}
\end{equation}
At every position, $D^{\mathrm{OPD}}_{t,i}>0$ means that the teacher assigns
the sampled token greater probability than the student, whereas
$D^{\mathrm{OPD}}_{t,i}<0$ expresses the opposite preference.  The signal is
dense because every generated position supplies one such comparison.

\paragraph{Regrouping by interaction turns.}
Equation~\ref{eq:app_sampled_opd} uses a single mean over the concatenated
generated sequence.  Partitioning its indices by turn and multiplying each
inner sum by $n_t/n_t$ gives the exact identity
\begin{equation}
\begin{split}
\frac{1}{N}\sum_t\sum_{i\in\mathcal{I}_t}D^{\mathrm{OPD}}_{t,i}
&=\sum_t\frac{n_t}{N}
\left(\frac{1}{n_t}\sum_{i\in\mathcal I_t}D^{\mathrm{OPD}}_{t,i}\right)\\
&=\sum_t
\underbrace{\frac{n_t}{N}}_{\text{budget prior}}
\underbrace{\left[\frac{1}{n_t}\sum_{i\in\mathcal I_t}
\log\frac{\pi_T(u_{t,i}\mid z_{t,i})}
{\pi_S(u_{t,i}\mid z_{t,i})}\right]}_{\text{turn-mean teacher signal (direction prior)}}.
\end{split}
\label{eq:app_decomposition}
\end{equation}
Let
\begin{equation}
b_t^{\mathrm{OPD}}=\frac{n_t}{N},
\qquad
\bar D_t^{\mathrm{OPD}}
=\frac{1}{n_t}\sum_{i\in\mathcal I_t}D^{\mathrm{OPD}}_{t,i}.
\label{eq:app_turn_statistics}
\end{equation}
Then the trajectory score is simply
$\sum_t b_t^{\mathrm{OPD}}\bar D_t^{\mathrm{OPD}}$.  The coefficient
$b_t^{\mathrm{OPD}}$ is the \emph{budget prior}: equal token weighting
allocates supervision to a turn in direct proportion to its length.  The
turn-mean signal $\bar D_t^{\mathrm{OPD}}$ is the
\emph{direction prior}: it determines whether the sampled action is moved
toward or away from the teacher's preference.

The timing of a search interaction makes the second prior explicit.  Let
$\mathcal F_t^{-}$ be the information available when $a_t$ is sampled and
$\mathcal F_t=\sigma(\mathcal F_t^{-},o_t)$ the information available after
tool execution.  Every prefix $z_{t,i}$ and every
$D^{\mathrm{OPD}}_{t,i}$ is $\mathcal F_t^{-}$-measurable.  In contrast,
the realized utility of the action is determined after $o_t$ arrives and is
$\mathcal F_t$-measurable.  Standard OPD therefore fixes both
$b_t^{\mathrm{OPD}}$ and $\bar D_t^{\mathrm{OPD}}$ before observing the
evidence produced by the search action.

IG-OPD replaces these two quantities separately.  ITC substitutes a
normalized posterior budget $q_t$ for $b_t^{\mathrm{OPD}}$, while IQC adds a
signed post-retrieval correction on the tokens that specify $a_t$.  The
token multiplier implementing the budget substitution follows directly
from Eq.~\ref{eq:app_decomposition}:
\begin{equation}
w_t=\frac{q_t}{b_t^{\mathrm{OPD}}}=\frac{Nq_t}{n_t},
\qquad
\frac{1}{N}\sum_t\sum_{i\in\mathcal I_t}
w_tD^{\mathrm{OPD}}_{t,i}
=\sum_t q_t\bar D_t^{\mathrm{OPD}}.
\label{eq:app_budget_substitution}
\end{equation}
Thus $\lambda=0$ yields $q_t=1/T$ and assigns every turn exactly the same
aggregate OPD budget, independently of its number of generated tokens.

\subsection{Policy-gradient Surrogate and Gradient Flow}
\label{app:gradient}

\paragraph{Score-function form of sampled OPD.}
The stop-gradient operation separates score construction from policy
optimization.  To see the resulting gradient, consider the negative KL at a
fixed prefix,
\begin{equation}
J_z(\theta)
=\mathbb E_{u\sim\pi_\theta(\cdot\mid z)}
\left[\log\pi_T(u\mid z)-\log\pi_\theta(u\mid z)\right].
\label{eq:app_fixed_prefix_objective}
\end{equation}
Differentiating the categorical expectation gives
\begin{align}
\nabla_\theta J_z(\theta)
&=\sum_u \nabla_\theta\pi_\theta(u\mid z)
   D_\theta^{\mathrm{OPD}}(u,z)
 -\sum_u\pi_\theta(u\mid z)\nabla_\theta\log\pi_\theta(u\mid z)
\nonumber\\
&=\mathbb E_{u\sim\pi_\theta}
\left[D_\theta^{\mathrm{OPD}}(u,z)
\nabla_\theta\log\pi_\theta(u\mid z)\right],
\label{eq:app_score_function_gradient}
\end{align}
because the second sum equals
$\sum_u\nabla_\theta\pi_\theta(u\mid z)=0$.  Let $\bar\theta$ denote the
fixed behavior parameters used to draw a token, and define the local
surrogate
\begin{equation}
J_z^{\mathrm{sur}}(\theta;\bar\theta)
=\mathbb E_{u\sim\pi_{\bar\theta}(\cdot\mid z)}
\left[\operatorname{sg}\!\left(D_{\bar\theta}^{\mathrm{OPD}}(u,z)\right)
\log\pi_\theta(u\mid z)\right].
\label{eq:app_detached_equivalence}
\end{equation}
Evaluated at the behavior policy,
$\nabla_\theta J_z^{\mathrm{sur}}(\theta;\bar\theta)
vert_{	heta=\bar\theta}
=\nabla_\theta J_z(\theta)
vert_{	heta=\bar\theta}$.  The detached value is
therefore the coefficient of a policy gradient; its effect enters through
$\nabla_\theta\log\pi_\theta$.

\paragraph{Calibrated per-token score.}
At update $k$, trajectories are sampled from the behavior policy
$\pi_{\theta_k}$.  The teacher and behavior-policy log probabilities define
\begin{equation}
D^{\mathrm{OPD}}_{t,i}
=\log\pi_T(u_{t,i}\mid z_{t,i})
-\log\pi_{\theta_k}(u_{t,i}\mid z_{t,i}).
\label{eq:app_behavior_gap}
\end{equation}
Let $\mathcal M_t\subseteq\mathcal I_t$ contain the action-critical tokens
of search action $a_t$, and let $m_t=|\mathcal M_t|$.  Combining ITC and
IQC gives
\begin{equation}
A_{t,i}
=\frac{Nq_t}{n_t}D^{\mathrm{OPD}}_{t,i}
+\mathbf 1[i\in\mathcal M_t]\frac{\alpha_k g_t}{m_t}.
\label{eq:app_calibrated_score}
\end{equation}
Both $q_t$ and $g_t$ are computed from the frozen teacher and are detached
along with $D^{\mathrm{OPD}}_{t,i}$.  For one trajectory, the corresponding
log-probability surrogate is
\begin{equation}
\mathcal{L}_{\mathrm{sur}}(\theta;\tau)
=-\frac{1}{N}\sum_t\sum_{i\in\mathcal{I}_t}
\operatorname{sg}(A_{t,i})\log\pi_\theta(u_{t,i}\mid z_{t,i}),
\label{eq:app_surrogate}
\end{equation}
where $\operatorname{sg}$ denotes stop-gradient.  At
$\theta=\theta_k$, substitution of Eq.~\ref{eq:app_calibrated_score} yields
\begin{equation}
\begin{split}
\nabla_\theta\mathcal{L}_{\mathrm{sur}}
={}&-\sum_t\frac{q_t}{n_t}\sum_{i\in\mathcal{I}_t}
D^{\mathrm{OPD}}_{t,i}\nabla_\theta\log\pi_\theta(u_{t,i}\mid z_{t,i})\\
&-\frac{\alpha_k}{N}\sum_{t:m_t>0}\frac{g_t}{m_t}
\sum_{i\in\mathcal{M}_t}\nabla_\theta\log\pi_\theta(u_{t,i}\mid z_{t,i}).
\end{split}
\label{eq:app_gradient}
\end{equation}
The first line assigns total OPD coefficient $q_t$ to turn $t$.  The second
line applies the signed evidence correction only to the tokens specifying
the search decision.  A positive $g_t$ increases their sampled
log-probabilities; a negative $g_t$ decreases them.  Retrieval, reference
answer scoring, and the computation of $q_t$ remain outside the autograd
graph, while the student parameters are updated through the two displayed
log-probability gradients.

\paragraph{Clipped on-policy implementation.}
During optimization, the actor uses the likelihood ratio
$\rho_{t,i}(\theta)=\pi_\theta(u_{t,i}\mid z_{t,i})/
\pi_{\theta_k}(u_{t,i}\mid z_{t,i})$ and the clipped surrogate
\begin{equation}
\mathcal{L}_{\mathrm{clip}}(\theta)
=-\frac{1}{Z}\sum_{\tau\in\mathcal{B}}\sum_{t,i\in\mathcal{I}_t}
\min\!\left\{
\rho_{t,i}\widehat A_{t,i},\;
\operatorname{clip}(\rho_{t,i},1-\delta,1+\delta)\widehat A_{t,i}
\right\},
\label{eq:app_clipped}
\end{equation}
where $\widehat A_{t,i}=\operatorname{sg}(A_{t,i})$, $\delta$ is the clipping
threshold, and $Z$ counts valid generated tokens in the reduction batch.
At $\theta=\theta_k$, $\rho_{t,i}=1$ and the gradient reduces to
Eq.~\ref{eq:app_gradient}.  Away from the behavior policy, clipping bounds
the change contributed by each sampled token.  Padding positions, tool
observations, and prompt tokens are removed by the generation mask before
the normalizer $Z$ is computed.

The two calibrations enter this surrogate through distinct algebraic paths:
$q_t$ rescales the dense OPD coefficient on all generated tokens in the
turn, whereas $g_t$ contributes an additive signed coefficient on
$\mathcal M_t$.  IQC is placed outside the ITC multiplier, so its aggregate
score is set by the realized information change rather than by the
length-derived or calibrated OPD budget.

\subsection{Power-posterior Interpretation}
\label{app:power_posterior}

\paragraph{Trajectory factorization.}
Let $P_{\mathrm{env}}$ denote the tool environment.  The trajectory
distribution induced by a policy $\pi$ factorizes as
\begin{equation}
p_\pi(\tau\mid x)
=\left[\prod_{t=1}^{T-1}
\pi(a_t\mid H_{t-1})
P_{\mathrm{env}}(o_t\mid H_{t-1},a_t)\right]
\pi(a_T\mid H_{T-1}).
\label{eq:app_trajectory_factorization}
\end{equation}
The student and teacher act in the same environment.  Hence the environment
transition terms cancel in their trajectory likelihood ratio:
\begin{equation}
\log\frac{p_T(\tau\mid x)}{p_S(\tau\mid x)}
=\sum_{t=1}^{T}
\log\frac{\pi_T(a_t\mid H_{t-1})}
{\pi_S(a_t\mid H_{t-1})}.
\label{eq:app_trajectory_ratio}
\end{equation}
This is the sequence-level counterpart of summing the token gaps in
Eq.~\ref{eq:app_token_gap}.

\paragraph{Evidence-conditioned target distribution.}
Write $\Psi_t=\exp(\Phi_t)$ and set $g_T=0$ for the final response.  The
post-retrieval gains telescope:
\begin{equation}
G(\tau)=\sum_{t=1}^{T-1}g_t
=\sum_{t=1}^{T-1}(\Phi_t-\Phi_{t-1})
=\Phi_{T-1}-\Phi_0.
\label{eq:app_gain_telescoping}
\end{equation}
At training step $k$, augment the sequence-level OPD score with this net
information change:
\begin{equation}
R_k(\tau)
=\log\frac{p_T(\tau\mid x)}{p_S(\tau\mid x)}
+\alpha_k\sum_{t<T}g_t
=\log\frac{p_T(\tau\mid x)\Psi_{T-1}(\tau)^{\alpha_k}}
{p_S(\tau\mid x)}-\alpha_k\Phi_0.
\label{eq:app_sequence_reward}
\end{equation}
The term $-\alpha_k\Phi_0$ depends on $(x,y^\star)$ but not on the sampled
trajectory.  Define the normalized evidence-conditioned teacher target
\begin{equation}
\widetilde p_{T,k}(\tau\mid x,y^\star)
=\frac{p_T(\tau\mid x)\Psi_{T-1}(\tau)^{\alpha_k}}
{\mathcal Z_k(x,y^\star)},
\qquad
\mathcal Z_k
=\mathbb{E}_{\tau\sim p_T}
\big[\Psi_{T-1}(\tau)^{\alpha_k}\big].
\label{eq:app_power_target}
\end{equation}
For fixed $(x,y^\star)$ and $\alpha_k$, both $\mathcal Z_k$ and $\Phi_0$
are independent of the student parameters.  Expanding the reverse KL gives
\begin{equation}
-\mathbb{E}_{\tau\sim p_S}[R_k(\tau)]
=D_{\mathrm{KL}}\!\left(p_S\,\|\,\widetilde p_{T,k}\right)
-\log\mathcal Z_k+\alpha_k\Phi_0.
\label{eq:app_power_kl}
\end{equation}
Minimizing the gain-augmented sequence objective is therefore equivalent,
up to constants independent of the student, to reverse-KL distillation from
$\widetilde p_{T,k}$.  The target retains the teacher trajectory prior and
tilts it by the post-retrieval answer potential.

The same result can be written as a product of local evidence updates:
\begin{equation}
\Psi_{T-1}(\tau)^{\alpha_k}
=\Psi_0^{\alpha_k}
\prod_{t=1}^{T-1}\exp(\alpha_k g_t).
\label{eq:app_incremental_power}
\end{equation}
Each returned observation contributes the multiplicative factor
$\exp(\alpha_k g_t)$: useful evidence raises the trajectory's target mass,
while misleading evidence lowers it.  For any two trajectories $\tau$ and
$\tau'$, their odds under the calibrated target satisfy
\begin{equation}
\frac{\widetilde p_{T,k}(\tau\mid x,y^\star)}
{\widetilde p_{T,k}(\tau'\mid x,y^\star)}
=\frac{p_T(\tau\mid x)}{p_T(\tau'\mid x)}
\left(
\frac{\Psi_{T-1}(\tau)}{\Psi_{T-1}(\tau')}
\right)^{\alpha_k}.
\label{eq:app_power_odds}
\end{equation}
At $\alpha_k=0$, the target is the teacher trajectory distribution.  As the
cosine schedule increases $\alpha_k$, post-retrieval answer confidence
contributes progressively more strongly to trajectory ranking.  The
token-level IQC term realizes this evidence tilt on the decision tokens of
each search action, while ITC supplies the complementary posterior turn
budget.

\subsection{Complete IG-OPD Algorithm}
\label{app:algorithm}

\paragraph{Reference-answer probes and masks.}
The frozen teacher scores only reference-answer tokens under the initial
context and each post-observation history, using teacher forcing rather
than generating another answer. Their mean log-likelihood yields the
potential and adjacent differences $g_t$ defined in the main text.
The response has $g_T=0$ but retains a positive ITC budget. For a search
action $a_t$, IQC masks select the parsed tool-name and supported
argument-value spans, mapped back to
generated-token positions; reasoning, wrapper tags, and observations are
excluded.

\begin{center}
\begin{minipage}{0.97\linewidth}
\newcounter{igopdalgorithm}
\refstepcounter{igopdalgorithm}
\label{alg:igopd}
\noindent\textbf{Algorithm \theigopdalgorithm\quad Information-Gain Calibrated OPD}
\smallskip\hrule\smallskip
\noindent\textbf{Input:} frozen teacher $\pi_T$, initialized student
$\pi_\theta$, training pairs $(x,y^\star)$, ITC strength $\lambda$, IQC
strength $\alpha_{\max}$, ramp horizon $K$, and clipping threshold $\delta$.
\begin{enumerate}
\setlength{\itemsep}{2pt}
\setlength{\parskip}{0pt}
\item At update $k$, sample a prompt batch and generate tool-interleaved
rollouts with the behavior policy $\pi_{\theta_k}$.  Record generated-token
indices $\mathcal I_t$, pre-action prefixes, tool observations, and parsed
action spans.
\item Score every generated token under $\pi_T$ and $\pi_{\theta_k}$ on the
same prefix, and compute $D^{\mathrm{OPD}}_{t,i}$ using
Eq.~\ref{eq:app_behavior_gap}.
\item For each rollout, evaluate $\Phi_t$ under
$H_0,H_1,\ldots,H_{T-1}$.  Set
$g_t=\Phi_t-\Phi_{t-1}$ for $t<T$ and $g_T=0$.
\item Compute the batch gain scale $\sigma_g$ and the normalized ITC budget
$q_t$ with Eq.~\ref{eq:itc_budget}.  Set the per-token turn multiplier
$w_t=Nq_t/n_t$.
\item Parse each search action to construct $\mathcal M_t$ and
$m_t=|\mathcal M_t|$.  Set
$\alpha_k=\frac{\alpha_{\max}}{2}
[1-\cos(\pi\min\{k/K,1\})]$.
\item Form the calibrated token score
$A_{t,i}=w_tD^{\mathrm{OPD}}_{t,i}
+\mathbf 1[i\in\mathcal M_t]\alpha_k g_t/m_t$,
using a zero IQC term whenever $m_t=0$.
\item Detach $A_{t,i}$, construct the token ratios against
$\pi_{\theta_k}$, and update the student by minimizing
Eq.~\ref{eq:app_clipped} over generated-token positions.
\item Replace the behavior policy with the updated student and repeat until
the prescribed training horizon is reached.
\end{enumerate}
\hrule
\end{minipage}
\end{center}

\clearpage

\section{Implementation Details}
\label{app:experimental_setup}

\subsection{Agent Environment and Teacher Training}
\label{app:tools}

\begin{table}[h]
\centering
\small
\caption{Tool environment shared by teacher training, student training,
and inference.}
\label{tab:app_tools}
\setlength{\tabcolsep}{4pt}
\renewcommand{\arraystretch}{1.08}
\begin{tabular}{@{}p{0.16\linewidth}p{0.23\linewidth}p{0.55\linewidth}@{}}
\toprule
\textbf{Family} & \textbf{Tool} &
\textbf{Backend and returned observation} \\
\midrule
Retrieval
& \texttt{text\_search}
& Serper organic search, Jina Reader page extraction, and
query-conditioned Qwen3-32B evidence summaries with titles and URLs. \\

& \texttt{image\_search}
& Serper Lens reverse-image retrieval over the original image or an
intermediate image identifier. \\

\midrule
Perception
& \texttt{layout\_parsing}
& PPStructureV3 with PP-OCRv5 and layout, table, formula, and chart
recognition; returns structured blocks, text, and Markdown. \\

& \texttt{crop}
& PIL region extraction; returns a new image identifier and region
metadata. \\

\midrule
Enhancement
& \texttt{perspective\_correct}
& OpenCV quadrilateral rectification and homography. \\

& \texttt{sharpen}
& Local unsharp masking for small or blurred visual content. \\

& \texttt{super\_resolution}
& EDSR $4\times$ upsampling for low-resolution regions. \\
\bottomrule
\end{tabular}
\end{table}

\paragraph{Action--observation environment.}
We implement multimodal search as a multi-turn action--observation process.
At turn $t$, the policy receives the original question and image together
with the complete interaction history. It then emits a reasoning block
followed by either one structured tool call or a final response. A tool call
is represented as a JSON object enclosed by \texttt{<tool\_call>} tags, and
the corresponding result is appended to the trajectory inside
\texttt{<observation>} tags. This creates an explicit sequence of policy
actions and environment observations that can be processed consistently
during data collection, SFT, on-policy distillation, and inference.

Each newly generated image is assigned a stable identifier
(\texttt{img\_1}, \texttt{img\_2}, \ldots). Subsequent tools can therefore
operate on a crop, rectified view, or enhanced version without serializing
the image into the textual context. The policy may execute at most one tool
per turn and up to 20 interaction turns per trajectory. Model-generated
reasoning, tool-call, and final-response tokens receive optimization masks.
System instructions, user content, images, and tool observations remain
visible as conditioning context but are excluded from the actor loss.

\paragraph{Search and retrieval stack.}
The textual retrieval tool first submits the query to the Serper Search API
and obtains the top organic results, with five results used by default. For
each result, the pipeline retains its title, URL, and search snippet and
uses Jina Reader to extract page-level content. The recovered passages are
then compressed by a query-conditioned Qwen3-32B summarizer exposed through
an OpenAI-compatible endpoint. The resulting observation preserves the
source URL and the evidence relevant to the current information need while
removing unrelated webpage content.

This retrieval pipeline allows the agent to refine its query over multiple
turns without repeatedly inserting entire webpages into the context. The
language parameter is passed to Serper, enabling multilingual queries and
cross-lingual retrieval through the same interface. Search observations
include the result title, source URL, and a focused evidence summary, which
allows later turns to verify, refine, or combine information from multiple
sources.

The image-retrieval tool uses the Serper Lens endpoint. It accepts either
the original image or an intermediate image identifier, performs a reverse
visual lookup, and returns candidate entities and visually related pages.
This is particularly useful when an object, artwork, landmark, product, or
document must first be identified before a factual text query can be
formed. The agent prompt instructs the policy to follow visual
identification with \texttt{text\_search} whenever the requested answer
depends on an externally verifiable fact.

\paragraph{Visual perception and enhancement stack.}
Document understanding is served locally through a FastAPI endpoint backed
by PaddleOCR PPStructureV3. The service combines PP-OCRv5 text detection and
recognition with dedicated document-layout, table, formula, and chart
modules. It returns recognized text, Markdown, reading-order information,
and typed layout blocks. Consequently, labels, tables, equations, captions,
and dense document regions can be incorporated into the trajectory as
structured observations rather than inferred solely from image pixels.

Lightweight visual transformations are executed locally. Cropping uses PIL
to extract a specified region. Perspective correction detects a
quadrilateral and applies an OpenCV homography. Sharpening uses unsharp
masking to improve the legibility of small or blurred text.
Super-resolution uses an EDSR $4\times$ model. These operations can be
composed into tool chains such as
\texttt{perspective\_correct}$\rightarrow$\texttt{crop}
$\rightarrow$\texttt{layout\_parsing}, while each intermediate image
remains addressable through its assigned identifier.

\paragraph{Training the teacher with custom GRPO rewards.}
Before distillation, we adapt OpenSearch-VL-8B to the tool environment with
GRPO. The reward jointly optimizes final-answer correctness, search
quality, structured execution, and search efficiency. For every sampled
trajectory $\tau$, we define
\begin{equation}
R_T(\tau)
=
r_{\mathrm{fmt}}\,e(s)
\left(
0.8r_{\mathrm{acc}}
+
0.2r_{\mathrm{query}}
\right),
\qquad
e(s)=
\begin{cases}
1, & s\leq5,\\
\max\{0.7,1-0.05(s-5)\}, & s>5,
\end{cases}
\label{eq:app_teacher_reward}
\end{equation}
where $r_{\mathrm{acc}}$ scores final-answer correctness and
$r_{\mathrm{query}}$ evaluates evidence relevance, query progression, and
the complementarity of visual and textual search. The variable $s$ counts
text and image searches.

The format term $r_{\mathrm{fmt}}$ is computed from the structural validity
of the generated steps. Well-formed reasoning, tool-call, observation, and
response blocks receive full credit. A valid direct-response fallback
receives partial credit, whereas malformed tool calls and execution-error
steps receive zero. The efficiency factor preserves the full task reward
for concise trajectories and gradually discounts redundant retrieval after
the fifth search without making search count the dominant objective.

For each prompt $x$, GRPO samples a group of $G=8$ trajectories and forms
the group-relative advantage
\begin{equation}
\widehat A_j
=
\frac{R_T(\tau_j)-\mu_x}{\sigma_x+\varepsilon},
\qquad
\mu_x
=
\frac{1}{G}\sum_{j=1}^{G}R_T(\tau_j),
\end{equation}
where $\mu_x$ and $\sigma_x$ are the mean and standard deviation within the
rollout group. The teacher actor is optimized using
\begin{equation}
\mathcal J_T(\theta)
=
\mathbb E_x
\left[
\frac{1}{G}
\sum_{j=1}^{G}
\sum_{i\in\mathcal I(\tau_j)}
\min
\left(
\rho_{j,i}(\theta)\widehat A_j,\,
\operatorname{clip}
\bigl(\rho_{j,i}(\theta),0.8,1.28\bigr)
\widehat A_j
\right)
\right],
\label{eq:app_teacher_grpo}
\end{equation}
where
\begin{equation}
\rho_{j,i}(\theta)
=
\frac{
\pi_\theta(u_{j,i}\mid z_{j,i})
}{
\pi_{\mathrm{old}}(u_{j,i}\mid z_{j,i})
},
\end{equation}
and $\mathcal I(\tau_j)$ contains only model-generated tokens. Masked token
scores are summed within each sequence and averaged across sequences.

Teacher training uses 64 prompts per update and eight responses per prompt,
corresponding to 512 sampled trajectories per update. We use a PPO
minibatch size of 16, an actor learning rate of $10^{-6}$, BF16 precision,
and one epoch on eight GPUs. Teacher rollouts use temperature 0.7 and
top-$p=1$. We use the resulting update-60 checkpoint as the distillation
teacher and freeze all teacher parameters throughout subsequent OPD and
IG-OPD training.

\subsection{OPD Data Construction}
\label{app:data}

\paragraph{Source mixture.}
The OPD corpus contains search-intensive multimodal questions rather than
precomputed teacher trajectories. Each record contains a unique training
identifier, a question, one or more image references, a reference answer
$y^\star$, and source metadata. The clean parent pool combines
Search-VL-RL-8K, VDR-Bench-2K, and the FVQA training split.

We sample 3,840 prompts without replacement using a fixed 70:15:15 source
ratio, as summarized in Table~\ref{tab:app_data}. Search-VL-RL provides
broad open-domain visual-search problems, while VDR-Bench and FVQA increase
the coverage of document-centric perception and fact-oriented retrieval.
The FVQA component is drawn exclusively from its training split.

\begin{table}[h]
\centering
\small
\caption{OPD prompt mixture shared by OPD, IG-OPD, and the on-policy
student baselines.}
\label{tab:app_data}
\setlength{\tabcolsep}{7pt}
\renewcommand{\arraystretch}{1.06}
\begin{tabular}{@{}lrrr@{}}
\toprule
\textbf{Source} & \textbf{Parent pool} &
\textbf{Selected} & \textbf{Share} \\
\midrule
Search-VL-RL-8K & 5,600 & 2,688 & 70\% \\
VDR-Bench-2K & 1,200 & 576 & 15\% \\
FVQA training split & 1,200 & 576 & 15\% \\
\midrule
Total & 8,000 & 3,840 & 100\% \\
\bottomrule
\end{tabular}
\end{table}

\paragraph{Schema validation and within-pool cleaning.}
We first convert the three sources into a common schema. Question text is
canonicalized by normalizing Unicode, letter case, punctuation, and
whitespace. Records with an empty question or answer, invalid image
references, duplicated training identifiers, or malformed fields are
removed before source-stratified sampling. 

% \paragraph{Training--evaluation decontamination.}
% To strengthen the validity of cross-benchmark comparisons, we filter the
% training pool against the complete question sets of SimpleVQA, LiveVQA,
% MMSearch, MMSearch+, FVQA, and BrowseComp-VL. Let $f(q)$ denote the
% normalized semantic embedding of question $q$, and let
% $\mathcal D_{\mathrm{eval}}$ be the union of these evaluation sets. For
% every training candidate $x$, we compute
% \begin{equation}
% s_{\max}(x)
% =
% \max_{e\in\mathcal D_{\mathrm{eval}}}
% \frac{
% f(q_x)^\top f(q_e)
% }{
% \lVert f(q_x)\rVert_2
% \lVert f(q_e)\rVert_2
% }.
% \label{eq:app_decontamination}
% \end{equation}
% Every training item satisfying
% \begin{equation}
% s_{\max}(x)>0.90
% \end{equation}
% is removed, in addition to normalized exact matches. The semantic pass
% captures paraphrases and near-duplicate formulations that survive lexical
% matching and substantially reduces question-level overlap between
% on-policy training and the reported test sets. Decontamination is performed before source-stratified sampling. Therefore,
% filtered items cannot be reintroduced when constructing the 3,840-example
% subset, and the final corpus still follows the target 70:15:15 proportions.
% This procedure produces a cleaner separation between training prompts and
% the six evaluation benchmarks while preserving the source diversity needed
% for open-domain search training.

\paragraph{Deterministic subset construction.}
Within each cleaned source partition, examples are sampled without
replacement using seed 42. The selected partitions are concatenated and
globally shuffled using the same seed. The builder verifies that all 3,840
output identifiers are unique and that every selected item has a matching
reference-answer entry. It then writes both JSONL and Parquet versions and
records a SHA-256 digest of the ordered identifier list.

\paragraph{Reference answers and reusable tool artifacts.}
Reference answers are stored in an identifier-keyed cache and serve two
roles. First, they provide the answer target used by outcome-reward
baselines. Second, IG-OPD inserts the same $y^\star$ into the frozen
teacher probes that produce $H_t$ and $H_t^+$ in
Appendix~\ref{app:method}. Dataset construction validates cache coverage for
every selected training identifier, eliminating online answer lookup from
the training loop.

Image paths point to canonical source assets rather than duplicated image
payloads. We additionally cache the initial-image visual-search response by
training identifier and image identity. Paired runs can therefore reuse
the same initial visual lookup, stabilizing this external component while
leaving query-dependent textual retrieval fully on-policy. All subsequent
queries and tool-selection decisions are generated by the current student
policy.

\subsection{Training Configurations}
\label{app:training}

\paragraph{SFT cold start.}
Both student scales start from Qwen3-VL-Instruct and undergo full-parameter
SFT before on-policy training. The vision tower and multimodal projector
remain trainable, allowing the model to learn visual inspection,
multi-turn reasoning, and the structured tool protocol jointly with
language generation. We use the \texttt{qwen3\_vl} conversation template, a maximum sequence
length of 24,000 tokens, maximum image and video budgets of 262,144 and
16,384 pixels, and BF16 precision. SFT uses a global batch size of 128, a
learning rate of $10^{-5}$, cosine decay with a 0.1 warmup fraction, two
epochs, gradient checkpointing, and ZeRO-3 over eight GPUs. The 2B and 4B
distillation runs are initialized from their corresponding update-154 SFT
checkpoints.

\paragraph{Shared on-policy training backbone.}
OPD and IG-OPD share the same student initialization, frozen teacher,
prompt ordering, rollout engine, optimizer, and distributed configuration.
Each update samples 64 prompts and one student trajectory per prompt. The
asynchronous SGLang engine executes up to 64 agent trajectories and 256
tool calls in parallel.

Rollouts use temperature 0.7, top-$p=1$, and disabled top-$k$ truncation.
We use tensor parallelism 2, a maximum prompt length of 20,000 tokens, a
maximum generated length of 70,000 tokens, and at most 20 agent turns. The
long response budget accommodates interleaved reasoning, tool calls,
retrieved observations, and the final response without truncating
multi-hop search trajectories.

The student actor is trained in BF16 with FSDP on eight GPUs. We use a
learning rate of $5\times10^{-6}$, cosine decay with a 0.1 warmup fraction,
a PPO minibatch size of 64, one actor epoch per rollout batch, clipping
radius 0.2, and dynamic token-based microbatching. Gradient checkpointing
and CPU parameter and optimizer offloading are used for long-context
updates. The teacher remains frozen and can be offloaded outside its
scoring windows.

As derived in Appendix~\ref{app:gradient}, optimization is applied only to
student-generated action tokens. Prompts and tool observations determine
the policy context but are not prediction targets. No entropy bonus or
additional KL reward is used in the student objective.

\begin{table}[h]
\centering
\small
\caption{Training configurations for the 2B and 4B students. OPD and
IG-OPD share the on-policy backbone; the final rows list the calibration
terms enabled by IG-OPD.}
\label{tab:app_training}
\setlength{\tabcolsep}{4.5pt}
\renewcommand{\arraystretch}{1.05}
\begin{tabular}{@{}p{0.17\linewidth}p{0.38\linewidth}
p{0.18\linewidth}p{0.18\linewidth}@{}}
\toprule
\textbf{Category} & \textbf{Parameter} &
\textbf{SFT} & \textbf{OPD / IG-OPD} \\
\midrule
Model
& Student scale & 2B / 4B & 2B / 4B \\
& Initialization & Qwen3-VL-Instruct & SFT update 154 \\
& Teacher & --- & OpenSearch-VL-8B, frozen \\
& Trainable parameters & Full model & Full student \\
& Precision & BF16 & BF16 \\

\midrule
Data
& Training instances & 9,843 trajectories & 3,840 prompts \\
& Global batch size & 128 trajectories & 64 prompts \\
& Image / video pixel budget & 262,144 / 16,384 & Input dependent \\

\midrule
Sequence
& Maximum sequence length & 24,000 & --- \\
& Maximum prompt / response length & --- & 20,000 / 70,000 \\
& Maximum agent turns & --- & 20 \\

\midrule
Optimization
& Learning rate & $10^{-5}$ & $5\times10^{-6}$ \\
& LR schedule / warmup fraction & Cosine / 0.1 & Cosine / 0.1 \\
& Training horizon & 2 epochs & 1 epoch \\
& PPO minibatch / actor epochs & --- & $64 / 1$ \\
& PPO clipping & --- & 0.2 \\
& Entropy / extra KL coefficient & --- & $0 / 0$ \\
& Gradient checkpointing & Yes & Yes \\

\midrule
Rollout
& Engine & --- & SGLang, asynchronous \\
& Responses per prompt & --- & 1 \\
& Temperature / top-$p$ & --- & $0.7 / 1.0$ \\
& Tensor parallelism & --- & 2 \\
& Parallel trajectories / tool calls & --- & $64 / 256$ \\

\midrule
IG-OPD
& ITC strength $\lambda$ & --- & 0.1 \\
& IQC strength / ramp horizon & --- & $10 / 60$ \\
& Gain / potential stabilizer & --- & $10^{-6} / 10^{-3}$ \\
& IQC placement & --- & Outside the ITC weight \\

\midrule
System
& GPUs & 8 & 8 \\
& Parallel strategy & ZeRO-3 & FSDP \\
& Checkpoint interval & End of epoch & 10 updates \\
\bottomrule
\end{tabular}
\end{table}

\clearpage

\section{Prompts}
\label{app:prompts}

\subsection{System Prompt}
\label{app:system_prompt}

The same system instruction is used for teacher data collection, on-policy
student rollouts, and inference. The prompt defines the action--observation
protocol, visual-inspection strategy, retrieval policy, and output format.
At runtime, the user message contains the tool schemas inside
\texttt{<tools>} tags, followed by the question and any attached image. The
following box preserves the operational instructions while retaining one
representative tool-use example.

\begin{tcolorbox}[
  breakable,
  colback=promptbg,
  colbacktitle=prompttitle,
  coltitle=white,
  colframe=prompttitle,
  title={Agent System Prompt},
  fonttitle=\bfseries\rmfamily,
  left=6pt,
  right=6pt,
  top=6pt,
  bottom=6pt,
  boxrule=0.6pt,
  arc=2pt
]
\begin{lstlisting}[
  basicstyle=\scriptsize\ttfamily,
  breaklines=true,
  columns=fullflexible,
  keepspaces=true,
  showstringspaces=false,
  frame=none,
  aboveskip=0pt,
  belowskip=0pt,
  xleftmargin=0pt,
  xrightmargin=0pt
]
You are an advanced Visual Investigation Agent. Answer user
questions with maximum precision by proactively using image-
processing and retrieval tools.

The attached image in this user message is named img_1. Tool
outputs that create new images are named img_2, img_3, and so on.
Always operate on the latest and clearest image version.

Core Philosophy: "Verify, Don't Guess"

1. Tool-First Mindset
If text is small, crop it. If text is blurry, sharpen it.
If the image is tilted, use perspective_correct. Do not rely
only on the internal visual encoder when a tool can provide a
clearer observation.

2. Chain Your Tools
A non-trivial visual query may require several operations. For
example:

perspective_correct -> crop -> layout_parsing

Continue until the relevant visual information is sufficiently
clear.

3. External Validation
Whenever the answer depends on facts that are not purely visible
in the image, call text_search. After image_search identifies an
entity, use text_search to retrieve and verify the factual detail.

4. Thinking Protocol
Before each action, emit a <think> block that:
- analyzes the request;
- assesses image quality, legibility, geometry, and target size;
- identifies the remaining information gap;
- selects one next action.

5. Tool Calling Format
Tool signatures are supplied inside <tools>...</tools>. Emit
exactly one JSON tool call inside <tool_call>...</tool_call>:

<tool_call>
{"name": "<function>", "arguments": {<arguments>}}
</tool_call>

Wait for the corresponding <observation> before selecting the
next action.

6. Result Interpretation
- Treat successful layout_parsing output as extracted document
  evidence.
- Treat image_search output as an initial entity-identification
  result.
- Treat text_search output as query-focused external evidence.
- Use titles and URLs in the retrieved observations when
  synthesizing the answer.

7. Output Rules
- Emit at most one tool call per turn.
- Never emit a tool call without a preceding <think> block.
- When sufficient evidence has been collected, place the final
  answer inside <response>...</response>.

Example:

<think>
The document in img_1 is skewed, so I will correct its
perspective before extracting the text.
</think>
<tool_call>
{"name":"perspective_correct",
 "arguments":{"image":"img_1"}}
</tool_call>
\end{lstlisting}
\end{tcolorbox}

For examples without an attached image, the image-reference paragraph is
replaced by an instruction stating that \texttt{img\_1} is unavailable.
The remaining reasoning, retrieval, and output rules are unchanged.

\subsection{Judge Prompt}
\label{app:judge_prompt}

We use \texttt{gpt-4o-2024-08-06} as the answer judge. The judge receives
the original question, the reference answer, and the complete final report.
It determines whether the report contains an answer consistent with the
reference and returns a binary decision together with a short explanation.

\definecolor{judgebg}{HTML}{F5F5F5}
\definecolor{judgetitle}{HTML}{37474F}

\begin{tcolorbox}[
  breakable,
  colback=judgebg,
  colbacktitle=judgetitle,
  coltitle=white,
  colframe=judgetitle,
  title={Answer Judge Prompt},
  fonttitle=\bfseries\rmfamily,
  left=6pt,
  right=6pt,
  top=6pt,
  bottom=6pt,
  boxrule=0.6pt,
  arc=2pt
]
\begin{lstlisting}[
  basicstyle=\scriptsize\ttfamily,
  breaklines=true,
  columns=fullflexible,
  keepspaces=true,
  showstringspaces=false,
  frame=none,
  aboveskip=0pt,
  belowskip=0pt,
  xleftmargin=0pt,
  xrightmargin=0pt
]
System:
You are a precise but high-recall answer judge.

User:
You are evaluating whether a deep-research report contains the
correct answer to the given question.

[Question]
{question}

[Correct Answer]
{correct_answer}

[Deep Research Report]
{response}

Determine whether the deep-research report contains the correct
answer anywhere in its content.

1. Read the complete report.
2. Locate the answer expressed by the report.
3. Compare it with the provided correct answer.
4. Accept equivalent entities, aliases, units, numerical forms,
   dates, and semantically equivalent wording.
5. The answer does not need to appear in a dedicated final-answer
   sentence.
6. Return "yes" when the report contains an answer consistent with
   the correct answer.
7. Return "no" when the answer is absent or contradicted by the
   report.

Output exactly:
correct: [yes/no]
reasoning: [brief explanation]
\end{lstlisting}
\end{tcolorbox}

The returned field \texttt{correct} is parsed as the binary correctness
label. Outputs that do not follow the required form are resubmitted with an
explicit formatting reminder.

\subsection{Tool Scheme}
\label{app:tool_scheme}

The tool registry is serialized into an OpenAI-style JSON array and inserted
into the user message inside \texttt{<tools>} tags. Each entry specifies the
function name, its role, the accepted argument schema, required fields, and
default values. The scheme below follows the tool registry used by the
on-policy rollout environment. The \texttt{web\_search} entry is a
compatibility alias of \texttt{text\_search}; both execute the same
Serper--Jina--Qwen retrieval pipeline. Here, we show part of our tool scheme.

\definecolor{toolbg}{HTML}{F3F7FC}
\definecolor{tooltitle}{HTML}{315D8A}

\begin{tcolorbox}[
  breakable,
  colback=toolbg,
  colbacktitle=tooltitle,
  coltitle=white,
  colframe=tooltitle,
  title={Tool Scheme},
  fonttitle=\bfseries\rmfamily,
  left=6pt,
  right=6pt,
  top=6pt,
  bottom=6pt,
  boxrule=0.6pt,
  arc=2pt
]
\begin{lstlisting}[
  basicstyle=\scriptsize\ttfamily,
  breaklines=true,
  columns=fullflexible,
  keepspaces=true,
  showstringspaces=false,
  frame=none,
  aboveskip=0pt,
  belowskip=0pt,
  xleftmargin=0pt,
  xrightmargin=0pt
]
<tools>
[
  {
    "type": "function",
    "function": {
      "name": "crop",
      "description":
        "Crop a specific region from an image.",
      "parameters": {
        "type": "object",
        "properties": {
          "image": {
            "type": "string",
            "description": "Image reference, e.g. img_1"
          },
          "x": {
            "type": "integer",
            "description": "Starting X coordinate"
          },
          "y": {
            "type": "integer",
            "description": "Starting Y coordinate"
          },
          "width": {
            "type": "integer",
            "description": "Width of the crop region"
          },
          "height": {
            "type": "integer",
            "description": "Height of the crop region"
          }
        },
        "required": ["image", "x", "y", "width", "height"]
      }
    }
  },

  {
    "type": "function",
    "function": {
      "name": "layout_parsing",
      "description":
        "Perform document layout parsing on an image to extract
         structured text.",
      "parameters": {
        "type": "object",
        "properties": {
          "image": {
            "type": "string",
            "description": "Image reference, e.g. img_1"
          },
          "file_path": {
            "type": "string",
            "description": "Absolute image path, when available"
          },
          "use_chart_recognition": {
            "type": "boolean",
            "default": false
          },
          "use_doc_orientation_classify": {
            "type": "boolean",
            "default": false
          }
        },
        "required": []
      }
    }
  },

  {
    "type": "function",
    "function": {
      "name": "image_search",
      "description":
        "Analyze an image using visual search to identify objects.",
      "parameters": {
        "type": "object",
        "properties": {
          "url": {
            "type": "string",
            "description":
              "Image reference, e.g. img_1, or a direct URL"
          }
        },
        "required": ["url"]
      }
    }
  },

  {
    "type": "function",
    "function": {
      "name": "text_search",
      "description":
        "Search the web and return query-relevant passages from
         the top-k results.",
      "parameters": {
        "type": "object",
        "properties": {
          "q": {
            "type": "string",
            "description": "Search query"
          },
          "query": {
            "type": "string",
            "description": "Alternative field for q"
          },
          "hl": {
            "type": "string",
            "default": "en"
          },
          "lang": {
            "type": "string",
            "description": "Alias for hl",
            "default": "en"
          },
          "top_k": {
            "type": "integer",
            "default": 5
          }
        },
        "required": []
      }
    }
  }
]
</tools>
\end{lstlisting}
\end{tcolorbox}

The rollout parser canonicalizes the aliases \texttt{query}$\rightarrow$
\texttt{q} and \texttt{lang}$\rightarrow$\texttt{hl}. Visual tools return
a new image identifier when they produce an image, while retrieval tools
return textual observations that remain in the trajectory context for
subsequent reasoning.

\clearpage
\section{Case Study}
\label{app:case_study}

\begin{figure}[h]
\centering
\includegraphics[width=\linewidth]
{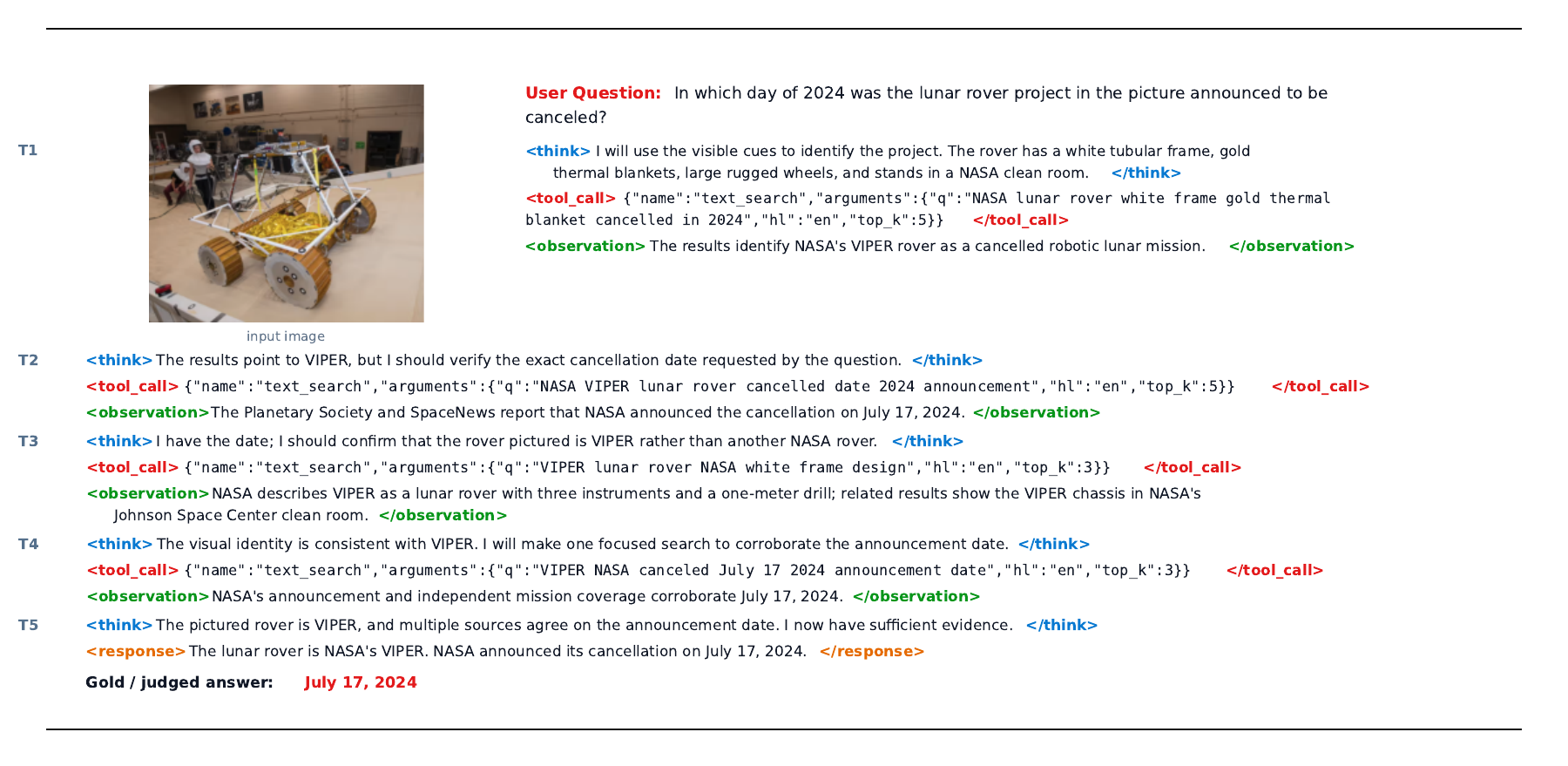}
\caption{A successful MMSearch trajectory. The student identifies the
pictured lunar rover as NASA's VIPER and corroborates its cancellation
date using multiple retrieval steps.}
\label{fig:case_viper}
\end{figure}

\begin{figure}[h]
\centering
\includegraphics[width=\linewidth]
{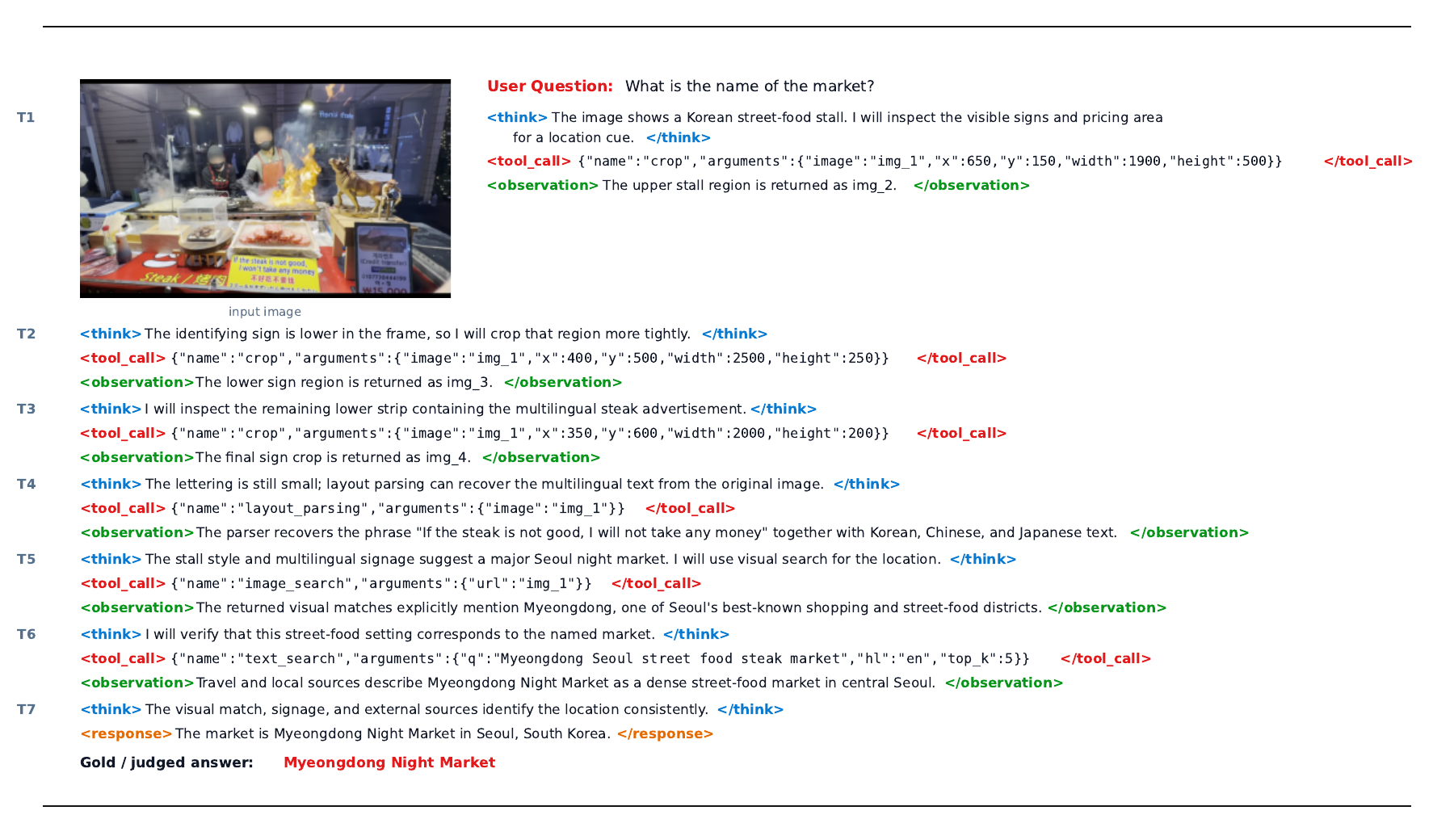}
\caption{A successful MMSearch+ trajectory. The student progressively
localizes the street-food scene through visual inspection, OCR evidence,
reverse-image retrieval, and textual verification.}
\label{fig:case_market}
\end{figure}

We present four successful trajectories produced by the 4B IG-OPD
student. The cases cover visual identification, document understanding,
image enhancement, and cross-entity multi-hop retrieval. Retrieved
passages are abridged for readability, while all retained tool calls and
their corresponding observations are shown. Figure~\ref{fig:case_viper} shows how the student identifies NASA's
VIPER rover from visual cues and progressively verifies the exact date
on which its cancellation was announced. Figure~\ref{fig:case_market} demonstrates a longer tool chain involving
region cropping, layout parsing, visual search, and textual verification.
The student combines multilingual signage and retrieved location evidence
to identify Myeongdong Night Market. 

\begin{figure}[h]
\centering
\includegraphics[width=\linewidth]
{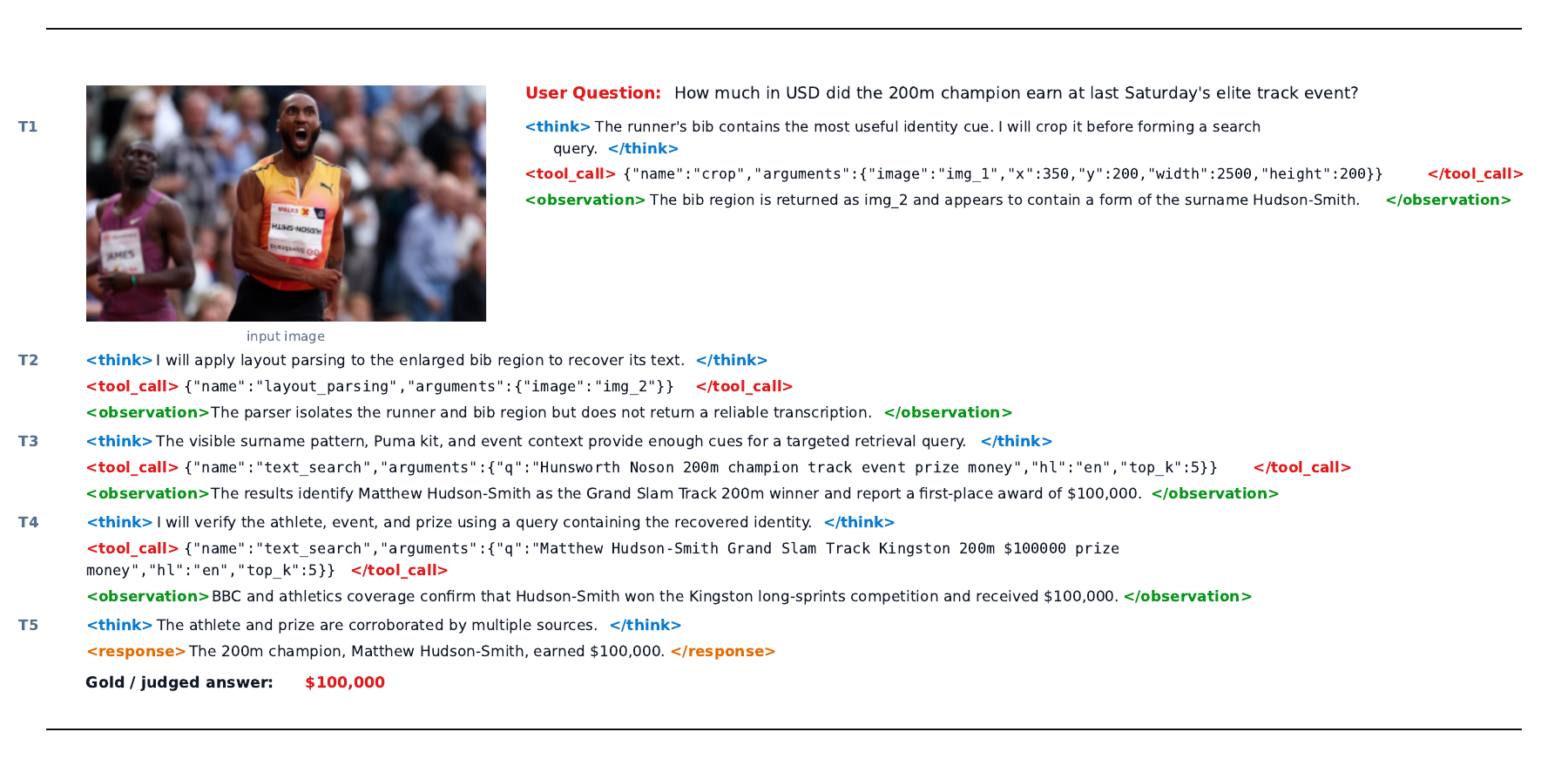}
\caption{A successful FVQA trajectory. The student localizes the
athlete's bib, identifies Matthew Hudson-Smith, and verifies the
\$100,000 prize through focused retrieval.}
\label{fig:case_track}
\end{figure}

\begin{figure}[h]
\centering
\includegraphics[width=\linewidth]
{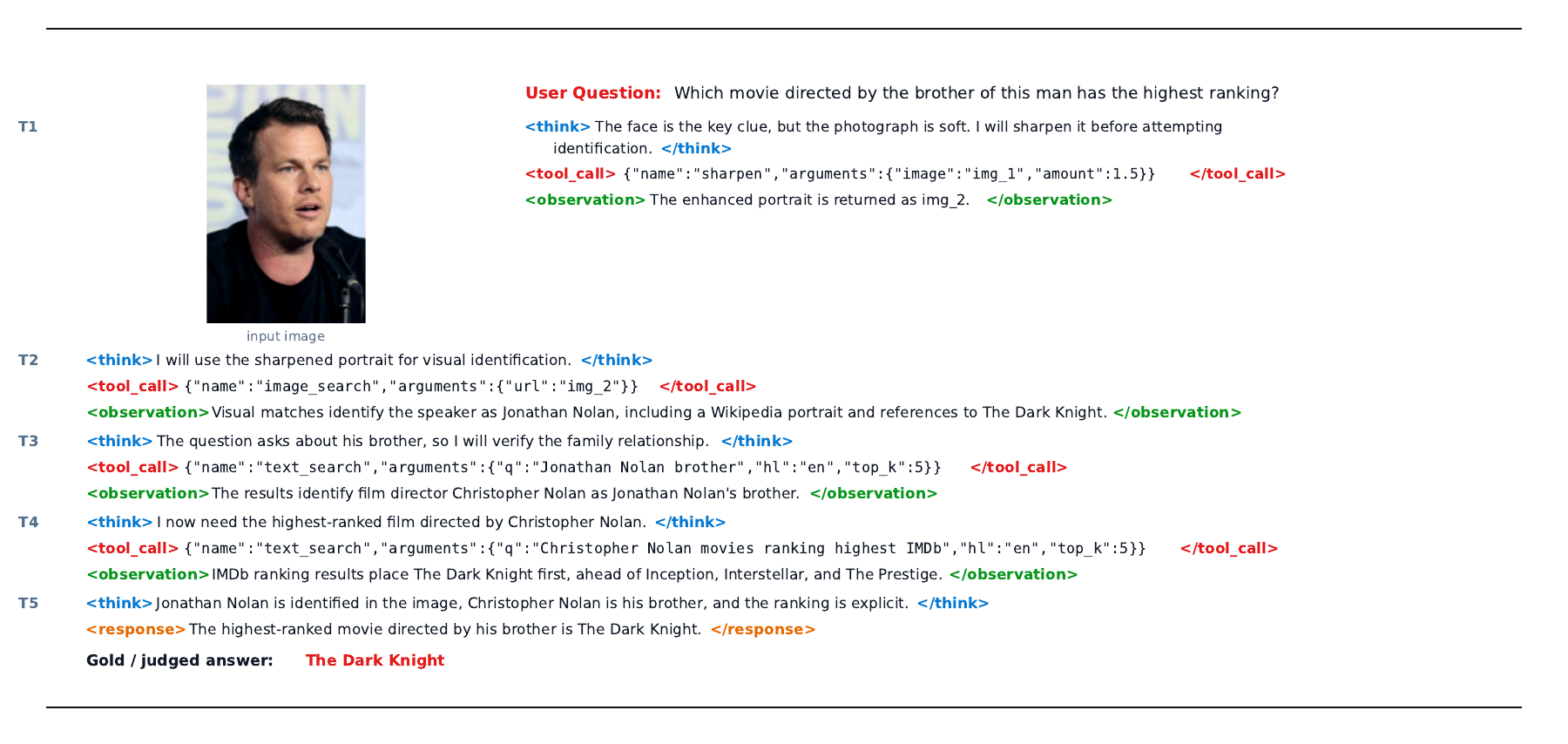}
\caption{A successful BrowseComp-VL trajectory involving image
enhancement, visual identification, relation retrieval, and ranking-based
factual verification.}
\label{fig:case_movie}
\end{figure}

Moreover, Figure~\ref{fig:case_track} illustrates how the student uses a small
textual cue on an athlete's bib to formulate a targeted search. It then
verifies both the athlete's identity and the corresponding prize amount
using independent sources. Figure~\ref{fig:case_movie} presents a cross-entity reasoning chain. The
student enhances the portrait, identifies Jonathan Nolan, retrieves his
relationship to Christopher Nolan, and then determines the highest-ranked
film directed by the latter.

\clearpage

\section{Ethics Statement}
\label{app:ethics}

This work advances the study of multimodal agents that acquire external
evidence through search and image-processing tools. Our experiments use
established research benchmarks and involve no human-subject studies or
collection of private personal data. By design, the experimental agent
operates in a read-only environment: it retrieves publicly accessible
information and processes benchmark images, while remaining strictly
confined to information retrieval and analysis without side effects on
external systems. We preserve source URLs for traceability and use
datasets and external services in accordance with their applicable
licenses and terms. These design choices enable transparent, reproducible,
and responsible evaluation of tool-augmented agents. We conduct and report
this work in accordance with the ICLR Code of Ethics.

\stopcontents[appendix]

\end{document}